\documentclass[11pt,letterpaper]{article}%
\pdfoutput=1
\usepackage[usenames,dvipsnames,svgnames,table]{xcolor}
\usepackage{color}
\usepackage{amsmath,amsfonts,amssymb}
\usepackage{verbatim}
\usepackage{relsize}
\usepackage{xargs}
\usepackage{epstopdf}
\usepackage{mathrsfs}
\usepackage{bm,bbm}
\usepackage{enumerate}
\usepackage[shortlabels]{enumitem}
\setlist{topsep=4pt,after=\vspace{\medskipamount},
         leftmargin=*,labelindent=8pt,labelsep=10pt
         }
\usepackage[square,numbers,merge,comma,sort&compress]{natbib} % bibliography
\makeatletter
\def\NAT@spacechar{\,}  % define space inside [1,\,2] ?
\makeatother

\usepackage{graphicx} % Required for including images
\graphicspath{{Figures/}} % Set the default folder for images

\usepackage[labelsep=colon]{caption}  % captions
\usepackage[labelsep=colon,aboveskip=10pt,belowskip=10pt]{subcaption}
\usepackage{jheppub}

\usepackage{xcolor}
\definecolor{Green}{rgb}{0.05, 0.45, 0.25}
\definecolor{RRed}{rgb}{0.65, 0.1, 0.5}

\usepackage{array,multirow,makecell,booktabs}  % tables
\newcolumntype{X}[2]{>{\centering\arraybackslash$}#1{#2\linewidth}<{$}}
\newcolumntype{R}[1]{>{\raggedleft\arraybackslash$}m{#1\linewidth}<{$}}
\newcolumntype{L}[1]{>{\raggedright\arraybackslash}m{#1\linewidth}}
\newcolumntype{M}[1]{>{\raggedright\arraybackslash$}m{#1\linewidth}<{$}}

\def\thinrule{\midrule[0.00001pt]}

\makeatletter
\renewcommand\mcell@classz{\@classx
   \@tempcnta \count@
   \prepnext@tok
   \@addtopreamble{%\mcell@mstyle
      \ifcase\@chnum
         \hfil
         \mcell@agape{\d@llarbegin\insert@column\d@llarend}\hfil \or
         \hskip1sp
         \mcell@agape{\d@llarbegin\insert@column\d@llarend}\hfil \or
         \hfil\hskip1sp
         \mcell@agape{\d@llarbegin \insert@column\d@llarend}\or
         \mcell@agape{$\vcenter
         \@startpbox{\@nextchar}\insert@column\@endpbox$}\or
         \mcell@agape{\vtop
         \@startpbox{\@nextchar}\insert@column\@endpbox}\or
         \mcell@agape{\vbox
         \@startpbox{\@nextchar}\insert@column\@endpbox}%
      \fi
      \global\let\mcell@left\relax\global\let\mcell@right\relax
    }\prepnext@tok}
\makeatother

\makeatletter
\newenvironment{subeqs}%
{\begingroup%
\setlength{\abovedisplayskip}{10pt plus 4pt minus 9pt}%
\setlength{\abovedisplayshortskip}{0pt plus 2pt minus 2pt}%
\setlength{\belowdisplayskip}{12pt plus 3pt minus 9pt}%
\setlength{\belowdisplayshortskip}{7pt plus 3pt minus 4pt}%
\begin{subequations}%
}%
{\end{subequations}\ignorespacesafterend
\endgroup}%
\makeatother

\makeatletter
\renewcommand{\thetable}{\arabic{table}\,}%{\color{\eqcol}\arabic{table}\,}
\renewcommand{\thefigure}{\arabic{figure}\,}%{\color{\eqcol}\arabic{figure}\,}
\renewcommand{\figurename}{Figure~\!\!\!}

\makeatother

\usepackage[breaklinks=true,backref=page]{hyperref}
\hypersetup{
    pdfpagemode={UseNone},
    pdfstartview={FitH},
    colorlinks=true,
    bookmarks=true,
    bookmarksnumbered=true,
    plainpages,
    a4paper,
    linktoc=page,
    citecolor=blue,
    filecolor=black,
    linkcolor=RRed,
    urlcolor=Green,
}
\renewcommand*{\backref}[1]{}
\renewcommand*{\backrefalt}[4]{%
\ifcase #1 %
\relax
\or
~{\small [\textsc{p.~\fns{\!#2}}]}
\else
~{\small [\textsc{p.~\fns{\!#2}}]}%
\fi}

\usepackage{footnotebackref}
\usepackage{hypernat}

\newcommand\I{\mathcal{I}}
\newcommand\R{\mathcal{R}}
\newcommandx{\II}[2][1=\Lambda,2=\Sigma,usedefault]{\mathcal{I}_\ms{#1#2}}
\newcommandx{\RR}[2][1=\Lambda,2=\Sigma,usedefault]{\mathcal{R}_\ms{#1#2}}
\newcommand{\V}{\mathcal{V}}
\newcommand{\zb}{\bar{z}}
\newcommand\fns{\footnotesize}

\newcommand\eD{\text{e}_\textsc{d}}

\newcommand\N{\mathcal{N}}
\newcommand\Ms{{\Scr M}}

\def\={~=~}
\def\<{~<~}
\def\>{~>~}
\def\*{{}^*}
\newcommand\eps{\epsilon}

\newcommand\LB{{\Scr L}_{\textsc{b}}}

\newcommand\EE{\textsc{e}}

\newcommand\MSK{\Ms_{\tts{SK}}}
\newcommand\Sact{\mathscr{S}}

\newcommand{\ms}{\mathsmaller}

\newcommandx{\tts}[1]{\text{\textsmaller{#1}}}
\newcommandx{\dm}[1][1=\mu,usedefault]{\partial_{#1}}
\newcommandx{\dmup}[1][1=\mu,usedefault]{\partial^{#1}}
\newcommandx{\Fd}[3][1=\Lambda,2=\mu,3=\nu,usedefault]{F^{\ms{#1}}_{#2#3}}
\newcommandx{\Fu}[3][1=\Lambda,2=\mu,3=\nu,usedefault]{{F^{\ms{#1}}}^{#2#3}}
\newcommandx\hodge{{}^\star}
\newcommandx{\LCTd}[4][1=\mu,2=\nu,3=\rho,4=\sigma,usedefault]{\eps_{#1#2#3#4}}
\newcommandx{\LCTu}[4][1=\mu,2=\nu,3=\rho,4=\sigma,usedefault]{\eps^{#1#2#3#4}}
\newcommand\z{z}

\newcommand\Sp{\mathrm{Sp}}
\newcommand\U{\mathrm{U}}

\newcommand\SO{\mathrm{SO}}
\newcommand\SL{\mathrm{SL}}
\newcommand\ISO{\mathrm{ISO}}

\DeclareFontFamily{OMS}{rsfs}{\skewchar\font'60}
\DeclareFontShape{OMS}{rsfs}{m}{n}{<-5>rsfs5 <5-7>rsfs7 <7->rsfs10 }{}
\DeclareSymbolFont{rsfs}{OMS}{rsfs}{m}{n}
\DeclareSymbolFontAlphabet{\Scr}{rsfs}

\title{Hairy Black Holes and Duality in an Extended Supergravity Model}
\author[a]{Andr\'{e}s Anabal\'{o}n}  %\note{...}}
\author[b]{, Dumitru Astefanesei}
\author[c]{, Antonio Gallerati}
\author[c]{, Mario Trigiante}
\author[]{.\medskip}

\affiliation[a]{Universidad Adolfo Ib\'{a}\~{n}ez; Dep.\ de Ciencias, Facultad de Artes Liberales - Vi\~{n}a del Mar, Chile.\medskip}
\affiliation[b]{Pontificia Universidad Cat\'{o}lica de Valpara\'{\i}so, Instituto de F\'{\i}sica - Valpara\'{\i}so, Chile.\medskip}
\affiliation[c]{Politecnico di Torino, Dipartimento DISAT - Torino, Italy.\\ \smallskip
Istituto Nazionale di Fisica Nucleare (INFN) - Sezione di Torino, Italy. \smallskip}
\bigskip

\abstract{We consider a $D=4$,\, $\N=2$ gauged supergravity with an electromagnetic Fayet-Iliopoulos term. We restrict to the uncharged, single dilaton consistent truncation and point out that the bulk Lagrangian is self-dual under electromagnetic duality. Within this truncation, we construct two families of exact hairy black hole solutions, which are asymptotically $AdS_4$.  When a duality transformation is applied on these solutions, they are mapped to two other inequivalent families of hairy black hole solutions. The mixed boundary conditions of the scalar field correspond to adding a triple-trace operator to the dual field theory action. We also show that this truncation contains all the consistent single dilaton truncations of gauged $\N=8$ supergravity with a possible $\omega$-deformation.}

\notoc

\begin{document}

\maketitle
\vspace*{\fill}

\newpage

\tableofcontents
\newpage

\section{Introduction}
The advent of gauge-gravity duality \cite{Maldacena:1997re} has produced a breakthrough in understanding the properties of strongly coupled field theories. In the context of string theory, where the AdS/CFT correspondence is best understood \cite{Gubser:1998bc, Witten:1998qj}, the physics in the bulk is described by theories of gravity coupled to various matter fields, which correspond to (consistent truncations of) gauged supergravities (see, e.g., \cite{Trigiante:2016mnt} and references therein). Since the AdS spacetime is not globally hyperbolic, the boundary conditions are crucial in order to obtain well-defined dynamics for a given field \cite{Ishibashi:2004wx}. For simplicity, one can, in principle, consider a bulk theory that includes just gravity and a single scalar field with a non-trivial potential. However, when the scalar field mass is in the Breitenlohner-Freedman (BF) window, there are infinitely many inequivalent boundary conditions which would imply the existence of a hairy soliton as a ground state. As it is possible to a priori pick the boundary which yields a desired ground state, the theories of this type
are referred to as `designer gravity' \cite{Hertog:2004ns}. According to the `holographic' dictionary, imposing mixed boundary conditions on the scalar field corresponds to perturbing boundary (UV critical point) conformal field theory by a relevant, irrelevant or marginal multi-trace deformation \cite{Witten:2001ua}.

Motivated by AdS/CFT duality, there has been extensive work on constructing exact neutral static black hole solutions with scalar hair \cite{Henneaux:2002wm, Martinez:2004nb, Hertog:2004dr, Anabalon:2012ta, Acena:2012mr, Anabalon:2013qua, Anabalon:2013sra, Acena:2013jya, Feng:2013tza, Wen:2015xea, Faedo:2015jqa, Fan:2015tua, Kichakova:2015nni} in designer gravity.  In general relativity, these solutions are important for clarifying different aspects of no-hair theorems \cite{Hertog:2006rr}, the role of ``scalar charges'' for black hole thermodynamics \cite{Hertog:2004bb, Anabalon:2014fla, Anabalon:2015xvl, Lu:2014maa}, and issues related to their stability \cite{Hertog:2005hm, Amsel:2007im, Faulkner:2010fh}. In a series of papers \cite{Anabalon:2012ta, Anabalon:2013qua, Anabalon:2013sra, Acena:2013jya}, by using a specific ansatz \cite{Anabalon:2012ta}, a new procedure was developed for obtaining exact regular hairy black hole solutions for a general scalar potential. At first sight, an issue related to this method is that the scalar potential is `engineered' and not obtained from physical considerations. However, intuitively, the condition for the existence of hairy black holes is that the self-interaction of the scalar field together with the gravitational interaction should combine such that the near-horizon hair does not collapse into the black hole while the far-region hair does not escape to infinity. In \cite{Nunez:1996xv}, it was shown in some concrete examples that, indeed, the hair should extend some way out from the event horizon and the degrees of freedom near the horizon are bound together with the ones that tend to be radiated away at infinity. This simple physical intuition hints to a direct connection between the integrability of the equations of motion and the form of the scalar potential \cite{Anabalon:2012ta, Anabalon:2013qua, Anabalon:2013sra, Acena:2013jya}. Not surprisingly, it was shown that, for some consistent truncations of supergravity theories for which exact neutral hairy black holes can be explicitly constructed \cite{Anabalon:2013eaa, Feng:2013tza, Lu:2014fpa, Faedo:2015jqa}, the corresponding scalar potentials are particular cases of this general potential. However, this raises an important question about the validity of exact hairy black hole generating technique and the embedding of its corresponding scalar potential in supergravity theories for which many physical aspects are under control.

In the recent years, after the surprising finding of a one-parameter family of $\SO(8)$ maximal four-dimensional supergravity theories \cite{DallAgata:2012mfj} with different physical properties with respect to the original $\SO(8)$ gauged model \cite{deWit:1982bul}, some progress was made towards the understanding of the vacuum structure \cite{Tarrio:2013qga, Anabalon:2013eaa, Borghese:2012qm, Borghese:2012zs, Borghese:2013dja, Dibitetto:2014sfa} and their dual field theory \cite{Pang:2015mra, Cremonini:2014gia}. Also, new domain wall and hairy black hole solutions were constructed \cite{Anabalon:2013eaa, Cremonini:2014gia}. Together with the original $\SO(8)$ model, other gauged supergravities have been extended by using dyonic embedding tensor \cite{Dall'Agata:2011aa,Dall'Agata:2012sx} (dyonic gaugings). These new gaugings feature a much richer vacuum structure and scalar field dynamics than their original counterparts.
Although the M-theory origin of those four-dimensional theories with dyonic gauging and semisimple gauge group remains yet to be elucidated, those with non-semisimple gauging have been uplifted to ten-dimensional type IIB or (massive) type IIA superstring theories \cite{Inverso:2016eet,Guarino:2015jca}. In particular, if the gauge group is chosen to be $\ISO(7)$, the concrete embedding in massive type IIA supergravity was obtained  in \cite{Guarino:2015jca} (see also \cite{Guarino:2015qaa, Guarino:2015vca, Ciceri:2016dmd}).

In this work, we focus on a particular class of models describing $\mathcal{N}=2$ supergravity coupled to a single vector multiplet in the presence of electric and magnetic Fayet-Iliopoulos (FI) terms. Analogous  theories were considered, with only electric FI terms, in \cite{Faedo:2015jqa}. After studying the AdS and dS vacua of the models for various choices of the parameters, we restrict ourselves to a consistent Einstein-dilaton truncation of the model and explicitly construct the scalar potential as a function of the dilaton only. Interestingly, after a redefinition of the parameters and a shift that makes the kinetic term of the dilaton canonically renormalized, the scalar potential matches (modulo a restriction on the parameters) the one of \cite{Anabalon:2012ta,Anabalon:2013qua, Anabalon:2013sra,Acena:2013jya}. We also discuss this truncation in the context of $\omega$-rotated ${\rm SO}(p,q)$-gauged maximal supergravities. This does not only close a gap on the physical significance of the scalar potential, but opens the possibility of constructing in a direct manner exact static regular hairy black hole solutions in supergravity.

From the perspective of gauge-gravity duality, our study can be motivated by the holographic relation between hairy solutions with mixed boundary conditions for the scalar field and the Renormalization Group (RG) flows of the dual field theory. The UV perturbation appears as boundary conditions on the supergravity fields at large radius. Relevant for phenomenology, the conformal invariance and supersymmetry can be broken by deformation of the gauge theory  by operators that produce an RG flow to one with a fewer symmetries. The boundary conditions can, in general, break conformal invariance \cite{Henneaux:2006hk}, but when the symmetry is preserved the geometry is AdS just asymptotically and becomes deformed due to the backreaction in the bulk. The conformal dimension of an operator can be related to the mass of AdS field to which it is isomorphic. By solving the equation of motion for a scalar field with $m^2 \geq 0$, one obtains two modes: the normalizable one is divergent in the interior and finite as one approaches the boundary of AdS spacetime and the other, which is referred to as non-normalizable,  is divergent at the boundary but finite in the interior. On one hand, exciting the normalizable mode significantly modifies the geometry of the bulk spacetime while preserving the asymptotic AdS behaviour. On the other hand, the non-normalizable modes are interpreted as boundary source currents for the operator dual to the scalar field \cite{Balasubramanian:1998sn, Balasubramanian:1998de}.

Therefore, an additional goal in this paper is to construct exact asymptotically AdS hairy black hole solutions in supergravity, study some of their generic features, and interpret them in the context of gauge-gravity duality. We obtain a general class of exact regular hairy black hole solutions that preserve the isometries of AdS in the boundary. These generalize to dyonic FI terms the solution found in \cite{Faedo:2015jqa}. The scalar field is tachyonic with mixed boundary conditions, but with a mass within the BF window, and so both modes are normalizable. From a holographic point of view,  the solutions correspond to adding triple trace deformations to the boundary action. We point out that the dyonic model has an unexpected symmetry that involves a transformation of the parameters of the theory. Remarkable enough, we show that the action of this symmetry on solutions is non-trivial. That provides a new solution generating technique in asymptotically AdS spacetimes, but the map between solutions modifies
the boundary conditions.\footnote{%
This is not generic in asymptotically AdS black holes since the solutions generating technique \cite{breitenlohner120four,Cvetic:1995kv,Cvetic:1996kv,gaiotto2007non,bergshoeff2009generating,bossard2009universal,fre2012integrability,andrianopoli2013extremal,andrianopoli2014extremal}, which is based on the global symmetry group of ungauged supergravities, can no longer be applied in the presence of a gauging, due to the non-trivial duality action on the embedding tensor.%
}

The remainder of the paper is organized as follows:\par
In Section \ref{sect2:theory}, after recalling the main facts about special geometry, we discuss the class of $\mathcal{N}=2$  models on which we shall focus in the present work. We discuss their general duality equivalence relations and vacua for various choices of the FI parameters.\par
In Section \ref{sect3:solutions}, we obtain a general Einstein-dilaton consistent truncation and, for any specific model, we show that there exist two families of exact hairy black hole solutions asymptoting a specific $AdS$ vacuum. These families are related to one another by an electric-magnetic duality transformation. The regularity condition and thermodynamics of these backgrounds are discussed in great detail.\par
In Section \ref{sect4:holography}, we provide a holographic interpretation of the solutions. We construct the regularized action in the gravity side, compare with the results of the dual field theory, and discuss the implications for our model.\par
In Section \ref{sect5:N8trunc} we show that, for suitable choices of the parameters, the class of Einstein-dilaton models considered here can be identified with truncations of $\omega$-rotated ${\rm SO}(p,q)$-gauged maximal supergravities \cite{DallAgata:2011aa,DallAgata:2012mfj,DallAgata:2012plb,DallAgata:2014tph,Inverso:2015viq}. This allows to embed the corresponding hairy black hole in those gauged maximal models.\par
Finally, we discuss some implications of our results, as well as possible future directions, in Section \ref{sect6:discuss}.

\section{The model}\label{sect2:theory}
In this section, after reviewing the main ideas behind special geometry in $\mathcal{N}=2$ theories, we shall focus on the model under consideration and discuss its main features.

\subsection{Brief review of special geometry}
Let us start recalling the relevant concepts about special K\"ahler manifolds, which is the class of target spaces spanned by the complex scalar fields in the vector multiplets of an $\mathcal{N}=2$ four-dimensional supergravity. This will also allow us to fix the notation.\par
Consider an $\mathcal{N}=2$ supergravity coupled to a number $n_v$ of vector multiplets and no hypermultiplet in the presence of FI terms. The theory describes $n_v+1$ vector fields $A^\Lambda_\mu$,\, $\Lambda=0,\dots, n_v$, and $n_s=n_v$ complex scalar fields $z^i$.
The general form of the bosonic Lagrangian is%
\footnote{%
Using the ``mostly minus'' convention and
\;$c=\hbar=1$. Moreover $\epsilon_{0123}=-\epsilon^{0123}=1$.
}:
\begin{equation}
\LB\=\frac{\eD}{8\pi G}\left(
-\frac{R}{2}
+g_{i\bar{\jmath}}\,\partial_\mu z^i\,\partial^\mu \bar{z}^{\bar{\jmath}}
+\frac{1}{4}\,\I_{\Lambda\Sigma}(z,\bar{z})\,F^\Lambda_{\mu\nu}\,F^{\Sigma\,\mu\nu}
+\frac{1}{8\,\eD}\,\R_{\Lambda\Sigma}(z,\bar{z})\,\eps^{\mu\nu\rho\sigma}\,F^\Lambda_{\mu\nu} \,F^{\Sigma}_{\rho\sigma}-V(z,\bar{z})\right)\,,
\label{boslagr}
\end{equation}
where \,$\eD=\sqrt{|\det(g_{\mu\nu})|}$\,. The $n_v+1$ vector field strengths are defined as usual:
\begin{align*}
F^\Lambda_{\mu\nu}\=\partial_\mu A^\Lambda_\nu-\partial_\nu A^\Lambda_\mu\;.
\end{align*}
The $n_s$ complex scalars $z^i$, $i=1,\dots, n_s$, couple to the vector fields in a non-minimal way through the real symmetric matrices $\I_{\Lambda\Sigma}(z,\bar{z})$, $\R_{\Lambda\Sigma}(z,\bar{z})$ and span a special K\"ahler manifold $\MSK$. The scalar potential originates from electric-magnetic FI terms. The presence of these terms amounts to gauging a ${\rm U}(1)$ symmetry of the corresponding ungauged model (with no FI terms) and implies minimal coupling of the vector fields to the fermions only.\par
For the general definition of special K\"ahler manifolds and its properties and the gauging procedure we refer the reader to the reviews \cite{Andrianopoli:1996cm,Trigiante:2016mnt,Gallerati:2016oyo}.
The geometry of $\MSK$ can be described in terms of a holomorphic section $\Omega^M(z^i)$ of the characteristic bundle defined over it, which is the product of a symplectic-bundle and a holomorphic line-bundle.
Denoting the components of $\Omega^M(z^i)$ as follows:
\begin{equation}
\Omega^M=\left(\begin{matrix}X^\Lambda\cr F_\Lambda\end{matrix}\right)\;,\qquad\quad\Lambda=0,\,\dots,n_v
\end{equation}
the K\"ahler potential and the metric have the following general form
\begin{equation}\label{Kom}
\begin{split}
\mathcal{K}(z,\bar{z})&~=-\log\left[i\,\overline{\Omega}^T\mathbb{C} \Omega\right]~=-\log\left[i\,\left(\overline{X}^\Lambda F_\Lambda-{X}^\Lambda \overline{F}_\Lambda\right)\right]\;,\\
g_{i\bar{\jmath}}~&=~\partial_i \partial_{\bar{\jmath}}\mathcal{K}\;,
\end{split}
\end{equation}
where $\mathbb{C}=(\mathbb{C}_{MN})$ is the $\Sp\big(2(n_v+1),\mathbb{R}\big)$-invariant antisymmetric matrix:
\begin{equation}
\mathbb{C}\=\left(\begin{matrix}{\bf 0} & {\bf 1}\cr -{\bf 1} & {\bf 0}\end{matrix}\right)\,.
\end{equation}
A change in the coordinate patch on the scalar manifold amounts to transforming $\Omega^M(z^i)$ by a corresponding constant $\Sp\big(2(n_v+1),\mathbb{R}\big)$ matrix besides multiplying it by a holomorphic function $\exp(f(z))$. The former leaves the K\"ahler potential invariant, as it is apparent from its manifestly symplectic invariant expression (\ref{Kom}), while the latter implies a corresponding K\"ahler transformation
\begin{equation}
\mathcal{K}(z,\bar{z}) \;\rightarrow\; \mathcal{K}(z,\bar{z})-f(z)-\bar{f}(\bar{z})\;.
\end{equation}
The choice of $\Omega^M(z^i)$, also fixes the symplectic frame (i.e.\ the basis of the symplectic fiber space) and thus the non-minimal couplings of the scalar fields to the vector field strengths in the Lagrangian. In the \emph{special coordinate frame} the lower components $F_\Lambda$ of the section are expressed as the gradient with respect to the upper entries $X^\Lambda$, of a characteristic prepotential function $F(X^\Lambda)$:
\begin{equation}
F_\Lambda\=\frac{\partial}{\partial X^\Lambda}F\;,
\end{equation}
where $F(X^\Lambda)$ is required to be a homogeneous function of degree two. In this frame the upper components $X^\Lambda(z^i)$, which are defined modulo multiplication times a holomorphic function, can be used as projective coordinates to describe the manifold and, in a local patch in which $X^0\neq 0$, we can identify the scalar fields with the ratios $z^i=X^i/X^0$.
We shall also use a covariantly holomorphic vector $\V^M=e^{\frac{\mathcal{K}}{2}}\,\Omega^M$, which is section of a $\U(1)$-line bundle satisfying the property:
\begin{equation}
D_{\bar{\imath}}\V^M~\equiv~ \left(\partial_{\bar{\imath}}-\frac{1}{2}\partial_{\bar{\imath}}\mathcal{K}\right)\V^M=0\,,
\end{equation}
$D_{i},\,D_{\bar{\imath}}$ being $\U(1)$-covariant derivatives. Under a K\"ahler transformation defined by a holomporphic function $f(z)$, this section transforms by a corresponding $\U(1)$-transformation:
\begin{equation}
\V^M \;\rightarrow\; e^{i\,\textrm{Im}(f)}\,\V^M\;.
\end{equation}
From its definition and eq.\ (\ref{Kom}), we find that $\V^M$ satisfies the condition $\V^T\mathbb{C} \overline{\V}=i$.  In particular the definition of this kind of manifold requires the section $\V^M$ to satisfy the additional properties:
\begin{equation}
\begin{split}
{D}_i U_j&\=i\,C_{ijk}\,g^{k\bar{k}}\,\overline{U}_{\bar{k}}\,,\\[\jot]
{D}_i\overline{U}_{\bar{\jmath}}&\=g_{i\bar{\jmath}}\,\overline{\V}\,,\\[\jot]
\V^T\mathbb{C}U_i&\=0\,,
\end{split}
\end{equation}
where we have defined
\begin{equation}
U_i^M\=\left(\begin{matrix}f_i^\Lambda\cr h_{\Lambda\,i}\end{matrix}\right)
    ~\equiv~ {D}_i \V^M \=\left(\partial_i+\frac{1}{2}\,\partial_i\mathcal{K}\right)\,\V^M\;,
\end{equation}
and $C_{ijk}$ is a characteristic covariantly holomorphic tensor which enters the expression of the Riemann tensor and defines the Pauli terms in the Lagrangian involving the gauginos.\par
The following identity holds:
\begin{equation}
U^{MN}~\equiv~ g^{i\bar{\jmath}}\,U_i^M \overline{U}_{\bar{\jmath}}^N~= -\frac{1}{2}\mathcal{M}^{MN}-\frac{i}{2}\,\mathbb{C}^{MN}
-\overline{\V}^M \V^N\,,\label{UMN}
\end{equation}
where $\mathcal{M}^{MN}$, and its inverse $\mathcal{M}_{MN}$, are symplectic, symmetric, negative definite matrices encoding the non-minimal couplings of the scalars $z^i$ to the vector fields in the Lagrangian. In particular $\mathcal{M}_{MN}$ has the following block-structure:
\begin{equation}  \label{M}
\mathcal{M}(\phi)\= (\mathcal{M}(\phi)_{MN}) ~\equiv~
\left(
\begin{matrix}
(\R\I^{-1}\R+\I)_{\Lambda\Sigma} &\;\;-(\R\I^{-1})_\Lambda{}^\Gamma \\ -(\I^{-1}\R)^\Delta{}_\Sigma & (\I^{-1})^{\Delta\Gamma} \\
\end{matrix}
\right)\;,
\end{equation}
and the matrices $\I,\,\R$ are those contracting the vector field strengths in (\ref{boslagr}). In terms of this matrix the couplings of the scalar fields to the vectors in the equations of motion can be written in a formally symplectic covariant form. To this end it is useful to introduce the symplectic vector of electric field strengths and their magnetic duals:
\begin{equation}
\mathbb{F}^M_{\mu\nu}\=
    \left(
    \begin{matrix}
    F^\Lambda_{\mu\nu}\cr G_{\Lambda\,\mu\nu}
    \end{matrix}
    \right)\,;
\end{equation}
where $G_{\Lambda\,\mu\nu}$ are the dual field strengths, functions of $F^\Lambda_{\mu\nu}$, of their Hodge duals and of the scalar fields:
\begin{equation}
G_{\Lambda\,\mu\nu}~\equiv
    -\epsilon_{\mu\nu\rho\sigma}\frac{\partial \mathscr{L}}{\partial F^\Lambda_{\rho\sigma}}\=
    \R_{\Lambda\Sigma}(z,\bar{z})\,F^\Sigma_{\mu\nu}-\I_{\Lambda\Sigma}(z,\bar{z})\,{}^*F^\Sigma_{\mu\nu}\;,
\end{equation}
and ${}^*F^\Lambda_{\mu\nu}\equiv \frac{\eD}{4}\,\epsilon_{\mu\nu\rho\sigma}\,F^{\Lambda\,\rho\sigma}$.\par
The equations of motion for the vector fields can be written in the following compact form:
\begin{equation}
d\mathbb{F}^M=0\;,\qquad\;
{}^*\mathbb{F}^M=-\mathbb{C}^{MN}\mathcal{M}_{NP}(z,\bar{z})\,\mathbb{F}^P\;,
\end{equation}
where we have used the exterior calculus notation with $\mathbb{F}^M\equiv \frac{1}{2}\, \mathbb{F}^M_{\mu\nu}\,dx^\mu\wedge dx^\nu$ and last equation is a formally symplectic covariant way of writing the definition of $G_{\Lambda\,\mu\nu}$.\par
We can deform the theory by introducing abelian electric-magnetic Fayet-Iliopoulos terms, defined by a constant symplectic vector $\theta_M$. This implies introducing a scalar potential $V(z,\bar{z})$ of the form:
\begin{equation}
V\=\left(U^{MN}-3\,\V^M\,\overline{\V}^N\right)\theta_M\,\theta_N~=
    -\frac{1}{2}\,\theta_M\mathcal{M}^{MN}\,\theta_N-4\,\V^M\,\overline{\V}^N\theta_M\,\theta_N\;.
\label{VFI}
\end{equation}
The reader can easily verify that the above potential can be expressed on terms of a complex superpotential
\begin{equation}
W\=V^M\,\theta_M\;,
\end{equation}
section of the $\U(1)$-bundle, as follows:
\begin{equation}
V\=g^{i\bar{\jmath}}D_iW\,D_{\bar{\jmath}}\overline{W}-3\,|W|^2\;.
\end{equation}
We can also define a real superpotential ${\Scr W}=|W|$ in terms of which the potential reads:
\begin{equation}
V\=4\,g^{i\bar{\jmath}}\partial_i {\Scr W}\,\partial_{\bar{\jmath}}{\Scr W}-3\,{\Scr W}^2\;.\label{realsuper}
\end{equation}
The scalar field equations can be written in the following form:
\begin{equation}
\nabla_\mu(\partial^\mu z^i)+\tilde{\Gamma}^i_{jk}\,\partial_\mu z^j\partial^\mu z^k - \frac{1}{8}\,g^{i\bar{\jmath}}\,\mathbb{F}^M_{\mu\nu}\,\partial_{\bar{\jmath}}
\mathcal{M}_{MN}(z,\bar{z})\,\mathbb{F}^{N\,\mu\nu}+g^{i\bar{\jmath}}\,\partial_{\bar{\jmath}}V\,=0\, ,
\end{equation}
where $\nabla_\mu$ is the covariant derivative, only containing the space-time Christoffel symbol and $\tilde{\Gamma}^i_{jk}$ is the connection on the K\"ahler manifold.\par
Finally, the Einstein equations read:
\begin{equation}
\mathcal{R}_{\mu\nu}\=2\,\partial_{(\mu} z^i\,\partial_{\nu)}\bar{z}^{\bar{\jmath}} \,g_{i\bar{\jmath}}+\frac{1}{2}\,\mathbb{F}^M_{\mu\rho}\,
\mathcal{M}_{MN}(z,\bar{z})\,\mathbb{F}^{N}{}_\nu{}^\rho- V\,g_{\mu\nu}\;.
\end{equation}
As pointed out earlier, the bundle structure defined on the scalar manifold allows to associate with a generic isometry transformation of the latter, a K\"ahler transformation and a constant symplectic transformation, belonging to the structure groups, acting on the symplectic section $V^M$ and its derivatives. From the explicit form of the bosonic field equations and of the scalar potential, it is apparent that an isometry transformation of the scalar manifold is formally an on-shell symmetry of the theory provided the corresponding symplectic transformation is made to act on the electric field strengths and their magnetic duals as well as on the FI terms:
\begin{align}
z^i \quad\rightarrow\quad z^{\prime\,i}(z^j)\;:\quad
\begin{cases}
&\V^M(z',\zb')=e^{i\,\textrm{Im}(f)}\,(S^{-1})_N{}^M\,\V^N(z,\zb)\;,\\[0.5ex] &\theta_M\;\rightarrow\; \theta^\prime_M=S_M{}^N\,\theta_N\;,\\[1ex]
&\mathbb{F}^M\;\rightarrow\;\mathbb{F}^{M\prime}=S^{-1}\,_N{}^M\,\mathbb{F}^N\;.
\end{cases}
\end{align}
This formal invariance, however, involving a non-trivial transformation of the parameters of the model, encoded in the FI terms, should be more appropriately regarded as an equivalence between different theories. \par
In what follows we shall focus on a specific $\mathcal{N}=2$ model with FI terms and restrict to a particular class of solutions which, as we shall see, will be mapped into one another by a global symmetry of the model in spite of these involving a non-trivial duality transformation on the FI terms.

\subsection{The model}
The theory under consideration is an $\mathcal{N}=2$ model with no hypermultiplets and a single vector multiplet ($n_v=1$) containing a complex scalar field $z$. The special K\"ahler manifold spanned by $z$  was considered in \cite{Faedo:2015jqa}.
The geometry of the spacial K\"ahler manifold is characterized by a prepotential of the form:
\begin{equation}
F(X^\Lambda)~=-\frac{i}{4}\,(X^0)^{n}\,(X^1)^{2-n}\;,
\end{equation}
the coordinate $z$ being identified with the ratio $X^1/X^0$. For two special values of $n$ the model is a consistent truncation of the so-called STU model. The latter is an $\mathcal{N}=2$ supergravity coupled to $n_v=3$ vector multiplets and characterized, in a suitable symplectic frame, by the prepotential:
\begin{equation}
F_\tts{STU}(X^\Lambda)~=-\frac{i}{4}\,\sqrt{X^0\,X^1\,X^2\,X^3}\;.
\end{equation}
The scalar manifold is symmetric of the form $\mathscr{M}_\tts{STU}=\big(\SL(2,\mathbb{R})/\SO(2)\big)^3$ and is spanned by the three complex scalars $z^i=X^i/X^0$, $i=1,2,3$. This model is in turn a consistent truncation of the maximal theory in four-dimensions.
For $n=1/2$ the model under consideration is the so-called $z^3$-model, whose manifold is $\SL(2,\mathbb{R})/\SO(2)$ and is embedded in that of the STU model through the identification $z^1=z^2=z^3=z$. For $n=1$ the special K\"ahler manifold has still the form $\SL(2,\mathbb{R})/\SO(2)$, which is now identified only with the first factor in $\mathscr{M}_\tts{STU}$, i.e.\ $z=z^1$. \par
After setting $X^0=1$, the holomorphic section $\Omega^M$ reads:
\begin{equation}
\Omega^M=
\left(\begin{matrix}
1  \cr  z  \cr -\frac{i}{4}\,n\,z^{2-n}  \cr -\frac{i}{4}\,(2-n)\,z^{1-n}
\end{matrix}\right)\;,
\end{equation}
and the K\"ahler potential $\mathcal{K}$ is computed using (\ref{VFI}) and has the expression
\begin{equation}
e^{-\mathcal{K}}\=\frac{1}{4}\,z^{1-n}\,\big(n\,z-(n-2)\,\bar{z}\big)+\text{c.c.}
\end{equation}
Writing \,$z=e^{\lambda\,\phi}+i\,\chi$\,, the truncation to the dilaton field $\phi$
\begin{equation}
\chi\=0\,,\qquad\; F^\Lambda_{\mu\nu}\=0\,,
\end{equation}
is consistent provided:
\begin{equation}
(2-n)\,\theta_1\, \theta_3 - n\,\theta_2\, \theta_4\=0\,,
\label{eq:construnc}
\end{equation}
which is clearly satisfied for the electric FI terms: $\theta_3=\theta_4=0$.\par
The metric restricted to the dilaton field  reads:
\begin{equation}
ds^2\=2\,g_{z\bar{z}}\, dz\,d\bar{z}\big\vert_{\chi=d\chi=0}\=\frac{1}{2}\lambda^2\,n\,(2-n)\,d\phi^2\;,
\label{modulimetric}
\end{equation}
and is positive provided $0<n<2$. Choosing
\begin{equation}
\lambda\=\sqrt{\frac{2}{n\,(2-n)}}\;,
\end{equation}
the kinetic term for $\phi$ is canonically normalized and the truncated action reads
\begin{equation}
\Sact~=-\frac{1}{8\pi G} \,\int_{M}d^{4}x\;\eD\,\left[\,\frac{R}{2}-\frac{1}{2}
\left(\partial\phi\right)^2+V(\phi)\,\right]\;.
\end{equation}
As a function of the dilaton only, the scalar potential has the following form:
\begin{equation}
\begin{split}
V\left(\phi\right)\=&-2\,e^{\lambda\,\phi\,(n-2)}\,\left(\frac{2\,n-1}{n}\,\theta_{1}^2
+4\,\theta_{1}\,\theta_{2}\;e^{\lambda\,\phi}
+\frac{2\,n-3}{n-2}\,\theta_{2}^2\;e^{2\,\lambda\,\phi}\right)-\\
&-\frac{1}{8}\;e^{-\lambda\,\phi\,(n-2)}\,\left[\left(2\,n-1\right)\,n\,\theta_{3}^{2}
-4\,\theta_{3}\,\theta_{4}\,n\,\left(n-2\right)\,e^{-\lambda\,\phi}
+\left(n-2\right)\,\left(2\,n-3\right)\,\theta_{4}^{2}\;e^{-2\,\lambda\,\phi}\right].
\label{eq:potdil}
\end{split}
\end{equation}
Note that the truncation is consistent at the level of scalar potential, but not of superpotential. More specifically if we consider the real superpotential ${\Scr W}$ in \eqref{realsuper}, one can check that,
when \eqref{eq:construnc} holds:
\begin{equation}
\left.\frac{\partial}{\partial\chi} {\Scr W}\right\vert_{\chi=0}\neq 0\,.\label{norealsuperpot}
\end{equation}
As a consequence of this, the potential of our dilaton truncation cannot, generically, be expressed in terms of a real superpotential.
We shall discuss below spherical solutions to this model with AdS invariant boundary conditions. If the truncation admitted a real superpotential, according to a conjecture of \cite{Hertog:2006rr}, the spherically symmetric solutions should belong to theories where the energy is unbounded from below. This conjecture does not apply to our case because we are mostly interested in the case where the superpotential is complex.

\subsubsection{Symmetries}
The potential is invariant under the simultaneous transformation
\begin{equation} \label{eq:invar1}
z\rightarrow \frac{1}{z}\;,\quad\; \theta_{1}\rightarrow\pm\,\frac{n}{4}\,\theta_{3}\;,\quad\;
\theta_{2}\rightarrow\pm\,\frac{2-n}{4}\,\theta_{4}\;,\quad\;
\theta_{3}\rightarrow\mp\,\frac{4}{n}\,\theta_{1}\;,\quad\;
\theta_{4}\rightarrow\mp\,\frac{4}{2-n}\,\theta_{2}\;,
\end{equation}
and this symmetry in the truncation to the dilaton field implies the transformation $\phi\rightarrow -\phi$.
%
%This transformation implies
%%
%\begin{equation}
%\theta_{1}\,\theta_{3}\rightarrow-\theta_{1}\,\theta_{3}\,,\qquad\quad
%\theta_{2}\,\theta_{4}\rightarrow-\theta_{2}\,\theta_{4}\,.%
%\end{equation}
%
Another invariance is obtained using
\begin{equation} \label{eq:invar2}
\theta_{1}\rightarrow\pm\,\frac{n}{4}\,\theta_{4}\;,\quad\;
\theta_{2}\rightarrow\pm\,\frac{2-n}{4}\,\theta_{3}\;,\quad\;
\theta_{3}\rightarrow\mp\,\frac{4}{n}\,\theta_{2}\;,\quad\;
\theta_{4}\rightarrow\mp\,\frac{4}{2-n}\,\theta_{1}\;,\quad\;
n \rightarrow 2-n\;.
\end{equation}

\subsubsection{Vacua}
Let us make the change of variable
\begin{equation} \label{eq:xvar}
\phi~\rightarrow~\frac{1}{\lambda}\,\log{x}\;,
\end{equation}
in the potential \eqref{eq:potdil}, using also the consistent truncation condition \eqref{eq:construnc}
\begin{equation}
\theta_2\=\frac{2-n}{n}\;\frac{\theta_1\,\theta_3}{\theta_4}\;.
\end{equation}
The resulting potential has the form:
\begin{equation} \label{eq:pot}
V_0(x)\=\frac{1}{8\,n^2\,\theta_4^2}\;x^{-n}\,\big(F(x)-G(x)\big)\;,
\end{equation}
where
\begin{equation} \label{eq:FG}
\begin{split}
F(x)&~=-16\,\theta_1^2\;x^{2n-2}\,\Big[(n-2)\,(2n-3)\,\theta_3^2\;x^2-4\,n\,(n-2)\,\theta_3\,\theta_4\;x+n\,(2\,n-1)\,\theta_4^2\Big]\;,\\[\jot]
G(x)&\=n^2\,\theta_4^2\,\Big[n\,(2n-1)\,\theta_3^2\;x^2+4\,n\,(2-n)\,\theta_3\,\theta_4\;x+(n-2)\,(2n-3)\,\theta_4^2\Big]\;,
\end{split}%
\end{equation}
with the condition $x>0$, to ensure the reality of the solution.\par\smallskip
Now, if we want to find the vacua of the theory, we have to solve the equation
\begin{equation}
\frac{dV_0(x)}{dx}\=0\;,
\end{equation}
where the derivative of the potential has the form:
\begin{equation}
\frac{dV_0(x)}{dx}\=\frac{n\,(n-2)\,\theta_3}{8\,x^{n+1}}\,\big(\theta_3\,x-\theta_4\big)\,\big(h(x)-g(x)\big)\;,
\end{equation}
with
\begin{equation} \label{eq:fg}
\begin{split}
h(x)&\=\left(\frac{4\,\theta_1}{n\,\theta_4}\right)^2\,\left[(3-2\,n)\,x^{2\,n-1}+(2\,n-1)\,\frac{\theta_4}{\theta_3}\,x^{2\,n-2}\right]\;,\\
g(x)&\=-(2\,n-1)\,x-(3-2\,n)\,\frac{\theta_4}{\theta_3}\;.
\end{split}
\end{equation}
We can easily see that one vacuum solution is given by
\begin{equation}
x_1\=\frac{\theta_4}{\theta_3} \qquad\quad \text{if} \quad \frac{\theta_4}{\theta_3}>0 \;,
\end{equation}
while other vacuum configurations come from the solutions of the equation
\begin{equation}\label{eq:zeroder}
h(x)\=g(x)\;.
\end{equation}
The above eq.~\eqref{eq:zeroder} has a variable number of solutions, depending on the range of values to which the number $n$ belongs.\par
Table \ref{tab:vacua} summarizes the number, types and mass of vacuum solutions for each possible interval of values of $n$.
%(see App.~\ref{app:vacua} for details).

\begin{table}[!h]
\captionsetup{aboveskip=20pt,belowskip=10pt}
\noindent\centering
\makegapedcells
\setcellgapes{17.5pt} % vertical space for cells
\begin{tabular}%{@{}ccc@{}}
{@{} R{0.15} X{m}{0.185} L{0.21} M{0.17} L{0.12} @{}} % @{} prevents white spaces at table end
\toprule
\midrule
%             &                               &                  &  \\
%\thinrule
0<n<\dfrac12  & \makecell[c]{%
                \theta_4/\theta_3<0
                \\[5\jot]
                \theta_4/\theta_3>0}  & \makecell[l]{%
                                        1 Anti de Sitter
                                        \\[5\jot]
                                        3 Anti de Sitter}   &\makecell[l]{%
                                                             m^2\,L=6\\[5\jot] m^2\,L=-2,6}
                                                                                   &  \makecell[l]{%
                                                                                   fig.~\subref{subfig:1AdS_case1}
                                                                                   \\[5\jot]
                                                                                   fig.~\subref{subfig:3AdS_case1}}\\
\thinrule
 n=\dfrac12   & \theta_4/\theta_3>0   &  1 Anti de Sitter    & m^2\,L=-2
                                                                              & fig.~\subref{subfig:1AdS_case2}\\
\thinrule
\dfrac12<n<1  & \makecell[c]{%
                \theta_4/\theta_3<0
                \\[5\jot]
                \theta_4/\theta_3>0}   & \makecell[l]{%
                                          1 de Sitter
                                          \\[5\jot]
                                          1 Anti de Sitter} & \makecell[l]{%
                                                               m^2\,L=6\\[5\jot] m^2\,L=-2}
                                                                             & \makecell[l]{%
                                                                              fig.~\subref{subfig:1dS}
                                                                              \\[5\jot]
                                                                              fig.~\subref{subfig:1AdS_case2}}\\
\thinrule
  n=1         & \theta_4/\theta_3>0    &  1 Anti de Sitter  &  m^2\,L=-2
                                                                             &
                                                                              fig.~\subref{subfig:1AdS_case2}\\
\thinrule
1<n<\dfrac32  & \makecell[c]{%
                \theta_4/\theta_3<0
                \\[5\jot]
                \theta_4/\theta_3>0}  & \makecell[l]{%
                                         1 de Sitter
                                          \\[5\jot]
                                         1 Anti de Sitter}   &\makecell[l]{%
                                                              m^2\,L=6\\[5\jot] m^2\,L=-2}
                                                                                  &
                                                                                  \makecell[l]{%
                                                                                   fig.~\subref{subfig:1dS}
                                                                                   \\[5\jot]
                                                                                   fig.~\subref{subfig:1AdS_case2}}\\
\thinrule
 n=\dfrac32   & \theta_4/\theta_3>0   &  1 Anti de Sitter    & m^2\,L=-2
                                                                              & fig.~\subref{subfig:1AdS_case2}
                                                                                                 \\
\thinrule
\dfrac32<n<2  & \makecell[c]{%
                \theta_4/\theta_3<0
                \\[5\jot]
                \theta_4/\theta_3>0}  & \makecell[l]{%
                                         1 Anti de Sitter
                                          \\[5\jot]
                                         3 Anti de Sitter}   &
                                                               \makecell[l]{%
                                                                m^2\,L=6\\[5\jot] m^2\,L=-2,6}
                                                                                    &
                                                                                     \makecell[l]{%
                                                                                      fig.~\subref{subfig:1AdS_case3}
                                                                                       \\[5\jot]
                                                                                      fig.~\subref{subfig:3AdS_case2}}
                                                                                                  \\
\midrule
\bottomrule
\end{tabular}
%\smallskip
\caption{Vacuum types}
\label{tab:vacua}
\end{table}

\begin{figure}[!h]
\captionsetup{aboveskip=15pt,belowskip=20pt}
\centering
\begin{subfigure}[b]{.4\linewidth}
\includegraphics[width=\linewidth]{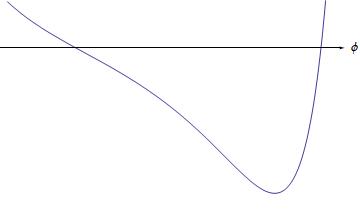}
\caption{}
\label{subfig:1AdS_case1}
\end{subfigure}
\qquad\qquad\qquad
\begin{subfigure}[b]{.4\linewidth}
\includegraphics[width=\linewidth]{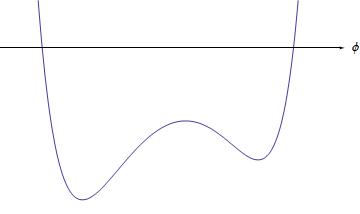}
\caption{}
\label{subfig:3AdS_case1}
\end{subfigure}
\\[15\jot]
\begin{subfigure}[b]{.4\linewidth}
\includegraphics[width=\linewidth]{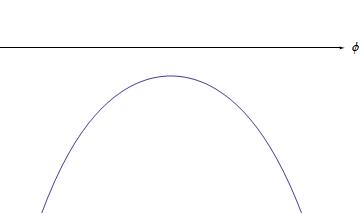}
\caption{}
\label{subfig:1AdS_case2}
\end{subfigure}
\qquad\qquad\qquad
\begin{subfigure}[b]{.4\linewidth}
\includegraphics[width=\linewidth]{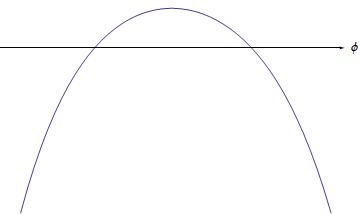}
\caption{}
\label{subfig:1dS}
\end{subfigure}
\\[15\jot]
\begin{subfigure}[b]{.4\linewidth}
\includegraphics[width=\linewidth]{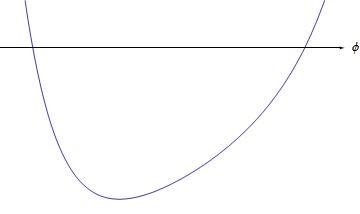}
\caption{}
\label{subfig:1AdS_case3}
\end{subfigure}
\qquad\qquad\qquad
\begin{subfigure}[b]{.4\linewidth}
\includegraphics[width=\linewidth]{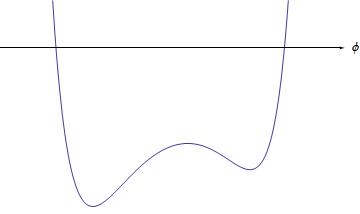}
\caption{}
\label{subfig:3AdS_case2}
\end{subfigure}
\caption{Vacua potential graphics}
\label{fig:vacua_graphs}
\end{figure}

\pagebreak

\section{Exact hairy black hole solutions}  \label{sect3:solutions}
In this section we construct exact hairy black hole solutions to the dilaton-truncation of the model that asymptote one of the previously discussed AdS vacua. After we redefine the parameters to obtain the dilaton potential of \cite{Anabalon:2012ta, Anabalon:2013qua, Anabalon:2013sra, Acena:2013jya}, we present the solutions, carefully check the existence of the horizon, and study their thermodynamics.

\subsection{A simplification of the potential}

To this end, it is convenient to redefine the parameters of the model.
First, we make a shift in the scalar field
\begin{equation}
\phi\=\varphi-\frac{2\nu}{\lambda\left(\nu+1\right)}\,\ln\left(\theta_{2}\,\xi\right)\;.\label{dilred}
\end{equation}
Then we redefine the parameters of the FI terms and the parameter $n$ as follows:
\begin{equation} \label{eq:newpar}
\theta_{1}=\frac{\nu+1}{\nu-1}\;\theta_{2}^{-\frac{\nu-1}{\nu+1}}\,\xi^{-\frac{2\nu}{\nu+1}}\,,\qquad
\theta_{3}=2\,\alpha\left(\xi\,\theta_{2}\right)^{\frac{\nu-1}{\nu+1}}\,s\,,\qquad
\theta_{4}=\frac{2\,\alpha}{\theta_{2}\,\xi\,s}\,,\qquad
n=1+\nu^{-1}\;,
\end{equation}
where we have introduced the new parameters $\alpha$, $\xi,$ $s$ and $\nu$. Next, it is convenient to express $\xi$ in terms of the AdS radius $L$:
\begin{equation}
\xi=\frac{2\,L\,\nu}{\nu-1}\,\frac{1}{\sqrt{1-\alpha^{2}\,L^{2}}}\,.
\end{equation}
Let us recall that the truncation to the dilaton is consistent provided equation \eqref{eq:construnc} is satisfied. In light of the new parametrization \eqref{eq:newpar}, this condition requires
\begin{equation}
(s^2-1)\,(\nu^2-1)\,\alpha\,\sqrt{1-L^2\,\alpha^2}\=0\;,
\end{equation}
which is solved, excluding values $n=0$ and $n=2$, either for pure electric FI terms ($\alpha=0$) or for $s=\pm1$. Since we are interested in dyonic FI terms, we shall restrict ourselves to the latter case.\par
The potential now reads
\begin{equation}
\begin{split}
V(\varphi)~=&-\frac{\alpha^2}{\nu^2}
\left[
\frac{(\nu-1)(\nu-2)}{2\,s^2}\,e^{-\varphi\,\ell\,(\nu+1)}
+ 2(\nu^{2}-1)\,e^{-\varphi\,\ell}
+ \frac{s^2}{2}\,(\nu+1)(\nu+2)\,e^{\varphi\,\ell\,(\nu-1)}
\right]+ \\
&+\left(\frac{\alpha^2-L^{-2}}{\nu^2}\right)
\left[
\frac{(\nu-1)(\nu-2)}{2}\,e^{\varphi\,\ell\,(\nu+1)}
+2(\nu^2-1)\,e^{\varphi\,\ell} + \frac{(\nu+1)(\nu+2)}{2}\,e^{-\varphi\,\ell\,(\nu-1)}
\right] ,
\end{split}
\end{equation}
where $\ell^{2}=\frac{2}{\nu^2-1}$ . We are left with three coupling
constants, as we have disposed of $\theta_{2}$ by the redefinition (\ref{dilred}) of the dilaton. The potential features a vacuum at $\varphi=0$ :
\begin{equation}
V(0)=-\frac{3}{L^{2}}\,,\qquad\quad
\left.\frac{dV(\varphi)}{d\varphi}\right\vert_{\varphi=0}\!\!=0\,,\qquad
\left.\frac{d^{2}V(\varphi)}{d\varphi^{2}}\right\vert_{\varphi=0}\!\!=-\frac{2}{L^{2}}\;.
\end{equation}
The scalar field mass is $m^2=-\frac{2}{L^{2}}$ and the theory is formally invariant under $\nu\leftrightarrow-\nu$, which corresponds to the equivalence \eqref{eq:invar2} and so, without loss of generality, we are going to consider $\nu>1$.\par
In what follows, we shall restrict ourselves to this vacuum and obtain two classes of black hole solutions. As we shall see below, the model features a duality invariance that acts on the FI terms as an electric-magnetic duality transformation, leaving the vacuum at $\varphi=0$ invariant.
This feature implies that the corresponding class of black hole solutions should also transform under the action of this duality.

\subsection{Symmetries and solutions of the truncation}
The theory is invariant under the change
\begin{equation} \label{eq:ST}%
\varphi\,\rightarrow\,-\varphi\,,\qquad
\alpha^{2}\,\rightarrow\,L^{-2}-\alpha^{2}\;,
\end{equation}
which is remnant of \eqref{eq:invar1} and  thus leaves invariant the products of the $\theta-$parameters
\begin{equation}
\theta_{1}\,\theta_{3}\,\rightarrow\,\theta_{1}\,\theta_{3}\;,\qquad\;
\theta_{2}\,\theta_{4}\,\rightarrow\,\theta_{2}\,\theta_{4}
\end{equation}
and corresponds to switching between electric and magnetic frames.

We start from the following general ansatz for the metric:
\begin{equation} \label{eq:AdSmetr}
ds^{2}\=
 \Upsilon(x)\,\left[f(x)\,dt^{2}-\frac{\eta^2}{f(x)}\,dx^2-d\Sigma_{k}\right]\;,
\end{equation}
\sloppy
where $x$ is the radial variable and $d\Sigma_{k}$ is the metric on the horizon. The latter is a two-dimensional manifold with constant curvature, fully characterized by its Ricci scalar  \,${R_{k}=2 k/L^2}$, with \,${k=0,-L^{-2},+L^{-2}}$\,. The parameter $\eta$ is the unique integration constant of the system non-trivially related to the mass. The engineering dimension of $\eta$ is $\left[\eta\right]=\left[M\right]^{-1}=\left[L\right]$.\par
The Einstein equations yield the following set of differential equations for the functions $f(x),\,\Upsilon(x)$:
\begin{align}
(\Upsilon\,f')'&=-2\,\eta^2\,k\,\Upsilon\,,\\
(\Upsilon\,f)''&=-2\,\eta^2\,\Upsilon^2\,V\,,\\
(\varphi')^2&=\frac{1}{2\Upsilon^2}\,\left(3\,(\Upsilon')^2-2\,\Upsilon''\Upsilon \right)\,,
\end{align}
where the prime symbol denotes the derivative with respect to $x$.
The equation of motion of the scalar field for this ansatz becomes
\begin{equation}
(\Upsilon\,f \varphi')'=\eta^2\,\Upsilon^2\,\partial_\varphi V\,.
\end{equation}
The relation between the above ansatz and that considered in \cite{Faedo:2015jqa}, characterized by the functions $X,\,Y,\,\psi$ and parameter $\beta$, can be obtained by the following redefinitions:
\begin{equation}
e^{2X}=\Upsilon\,f\;,\qquad
e^{2Y}=\Upsilon\;,\qquad
dr=\Upsilon\,\eta\,dx\;,\qquad
\beta=\frac{L^2\,(\nu^2-1)}{4\nu\eta^2}\;.
\end{equation}
We find two families of solutions, asymptoting the AdS vacuum at $\varphi=0$, which we are going to discuss in the following. They are related to each other through the action of the duality \eqref{eq:ST}.

\subsubsection{Family 1}
The functions $\varphi(x),\,\Upsilon(x),\,f(x)$ read
\begin{equation} \label{eq:AdSfunc}
\begin{split}
\varphi(x)&~=-\ell^{-1}\,\ln(x)\;, \qquad\quad\;
\Upsilon(x)~=\frac{L^2\nu^2\,x^{\nu-1}}{\eta^2\,(x^\nu-1)^2}\;, \\[3.5\jot]
f(x)&\=\frac{x^{2-\nu}\,(x^\nu-1)^2\,\eta^2\,k}{\nu^2} +
     1+\alpha^{2}L^2\,\left[-1+\frac{x^2}{\nu^2}\,\left((\nu+2)\,x^{-\nu}-(\nu-2)\,x^\nu+\nu^2-4\right)\right]\;.
\end{split}
\end{equation}
This family of solutions does not have a horizon when the magnetic gauging vanishes
($\alpha=0$).  \par
The asymptotic region is located at the pole of order 2 of the conformal factor, namely $x=1$. The geometry and scalar field are singular at $x=0$ and $x=\infty$. Therefore, the configuration contains two disjoint geometries given by $x\in(1,\infty)$ or $x\in(0,1)$. These geometries have an invariant characterization in terms of the scalar field:
\begin{equation}
\begin{split}
x\in(0,1) \quad\Longrightarrow\quad \varphi\geq0\;, \\
x\in(1,\infty) \quad\Longrightarrow\quad \varphi\leq0\;,
\end{split}
\end{equation}
and we shall call them the positive and the negative branch, accordingly.

\paragraph{Boundary conditions and mass:} To compare with the AdS canonical coordinates, let us consider the following fall off:
\begin{equation}
\Upsilon(x)\= \frac{r^2}{L^2} + O\left(r^{-2}\right)\;.
\end{equation}
Comparing with the specific expression of our solution, we found that the asymptotic change of coordinates that provides the right behaviour is
\begin{equation} \label{CC}%
x\= 1 \pm \left(\frac{L^2}{r\,\eta}+L^6\,\frac{1-\nu^2}{24\,\left(r\,\eta\right)^3}\right)  +L^8\,\frac{\nu^2-1}{24\,\left(r\,\eta\right)^4}\;,
\end{equation}
where we take $\eta>0$ and the $\pm$ sign depends on whether one takes the negative ($+$) or the
positive ($-$) branch. Accordingly, the scalar field behaves at the boundary as
\begin{equation}\label{fall off}
\varphi\=L^2\,\frac{\varphi_0}{r}+L^4\,\frac{\varphi_1}{r^2}+O\left(r^{-3}\right)\=
    \mp L^2\,\frac{1}{\ell\,\eta\,r}+L^4\,\frac{1}{2\,\ell\,\eta^2\,r^2}+ O\left(r^{-3}\right)\;,
\end{equation}
where we have normalized $\varphi_0$ and $\varphi_1$ to match their conformal and engineering dimension. In the canonical coordinates, we can now easily read off the coefficients of the leading and subleading terms in the scalar boundary expansion
\begin{equation}
\label{f1}
\varphi_0\=\mp\frac{1}{\ell\,\eta}\;, \qquad
\varphi_1\=\frac{\ell}{2}\,\varphi_{0}^2\;,
\end{equation}
which correspond to boundary conditions that are AdS invariant. In this case, since the boundary conformal symmetry is preserved, the black hole mass \cite{Anabalon:2014fla} can be read off from the metric \eqref{eq:AdSmetr}.  To do that, let us consider the asymptotic expansion of the other metric components:
\begin{align}
g_{tt}\= &\frac{r^2}{L^2}+k\,L^2-\frac{\mu_1 L^4}{r}
    +O\left(r^{-2}\right)\;,\\[2\jot]
g_{rr} ~= &-\frac{L^2}{r^2}-L^6\,\frac{k+\varphi_0^2/2}{r^4}
          +O\left(r^{-5}\right)\;,
\end{align}
where
\begin{equation}
\mu_1=\pm\left(\frac{\nu^2-4}{3\,\eta^3}\,\alpha^2\,L^2-\frac{k}{\eta}\right)\;,
\end{equation}
and we have taken $\eta>0$. Thus, the black hole mass is
\begin{equation}
M=L^4\,\frac{\mu_{1}\,\sigma_k}{8\pi G}\;,
\end{equation}
where, for convenience, $\sigma_k=L^{-2}\int d\Sigma_k$ is defined to be dimensionless. Concretely the values are \,$\sigma_1=4\pi$ and \,$\sigma_{-1}=8\pi (g-1)$,\, where $g \geq 2$ is the genus of a compact negative constant curvature manifold. When $\nu<2$, only the positive branch exists and the mass is positive, but for $\nu>2$ both branches can have positive mass. The case with $\nu=2$ should be treated separately. In the boundary, after discarding the conformal factor, the dual field theory lives
on a manifold of radius $L$ and, consequently, the energy density is
\begin{equation}
\label{rho}
\rho\=\frac{L^2}{8\pi G}\mu_{1}\;.
\end{equation}

\paragraph{Location of the horizon:} Let us, for the sake of simplicity, start focusing on the case of spherical topology ($k=L^{-2}$). The boundary is located at $x=1$ where the conformal factor blows up and the dilaton value vanishes. One way to check if there exist a horizon is to investigate if $f(x)$ changes its sign (in the boundary, $f(x=1)\,=\,1$). By verifying its expression close to the singularities, we get
\begin{equation}%
\begin{aligned}
& x<1  & \quad & \nu<2    & \quad &
                          f(x=0)=1-\alpha^2 L^2\; ,\\[2\jot]
& x<1  & \quad & \nu>2    & \quad &
                          f(x=0)=\big[\eta^2\,k+\alpha^2 L^2 \,(\nu+2)\big]\,\frac{ x^{2-\nu}}{\nu^2}\;,\\[2\jot]
& x>1  & \quad & \nu\neq2 & \quad &
                          f(x=+\infty)=\big[\eta^2\,k-\alpha^2 L^2\,(\nu-2)\big]\, \frac{ x^{2+\nu}}{\nu^2}\;.
\end{aligned}
\end{equation}
From this analysis, we conclude that there are black holes only for $\varphi_{0}<0$ and
\begin{equation} \label{eq:bhcond}
\eta^2\,k-\alpha^2 L^2\,(\nu-2)\<0\;,
\end{equation}
which requires that $\nu>2$. The existence of a horizon implies that the energy density is positive:
\begin{equation} \label{bound}
\rho_\text{sph}=\frac{L^2}{8 \pi G} \left(\frac{\nu^2-4}{3\,\eta^3}\,\alpha^2 L^2-\frac{1}{\eta L^2}\right)
\;>\;\frac{1}{8 \pi G} \frac{\nu-1}{3\,\eta}\;>\;\frac{1}{24 \pi G  L^2} \frac{\nu-1}{\left\vert\alpha \right\vert \sqrt{\nu-2}}\;.
\end{equation}
Hence, we have shown that on our configurations regularity implies positivity for the energy. The energy is bounded from below uniquely in terms of the parameters of the action. This has a form of a positive energy theorem without using spinorial techniques. Indeed, this analysis needs to be complemented by showing that there is no additional branch of solutions for this set of boundary conditions, but that lies outside the scope of our paper.

\paragraph{A detailed study of regularity:} Let us consider the different values of the parameters for which regular solutions in the family under consideration exist.
\begin{enumerate}[label=$\boldsymbol{\diamond}$]
\item{$\boldsymbol{k=0\,.}$\quad
    The function $f(x)$ has only one zero for $\nu>2$ in the region $x>1$, which defines the position of the horizon. Positivity of the energy requires $\eta>0$. For $\nu\le 2$ there is no horizon.}
\item{$\boldsymbol{k=L^{-2}\,.}$\quad
    Let us define the parameter $a=2+\frac{k\,\eta^2}{\alpha^2\,L^2}$\,.
    We have the following two cases:
    \begin{enumerate}[label=$\ms{\circ}$,topsep=-0.1em,itemsep=0.5em]
    \item{$\nu >a$\,:\quad
        the function $f(x)$ has only one zero in the region $x>1$. Positivity of the energy requires $\eta>0$. This implies the existence of a regular black hole;}
    \item{$\nu\le a$\,:\quad
         there is no horizon.}
    \end{enumerate}
     }
\item{$\boldsymbol{k=-L^{-2}\,.}$\quad
    With reference to the parameter $a$ previously defined, we distinguish among the following cases:
    \begin{enumerate}[label=$\ms{\circ}$,topsep=0em,itemsep=0.75em]
    \item{$\nu >{\rm max}(|a|,\,2)$\,:\\[\jot]
        the function $f(x)$ has only one zero in the region $x>1$. Positivity of the energy requires $\eta>0$;}
    \item{$|a|<\nu< 2$\,:\\[\jot]
         we have a zero of $f(x)$ for $x>1$ and the mass is positive for suitable values of $\eta$;\footnote{The condition on $\eta$ is $\frac{\sqrt{4-\nu^2}}{3}\,L^2\,|\alpha|< \eta<L^2\,\sqrt{2+\nu}\,|\alpha|$;}}
     \item{$-a<\nu<a$\,:\\[\jot]
        $f(x)$ can have no zeros or two zeros (which may coincide), for $x>1$. In the former case there is no regular solution. The mass is always negative;}
     \item{$\nu<-a$ \,and\, $\nu<2$\,:\\[\jot]
        $f(x)$ can have either two zeros for $x<1$ and one zero for $x>1$, or only one zero for $x>1$. The mass is positive for $x>1$, $\eta>L^2\,\sqrt{2+\nu}\,|\alpha|$;}
    \item{$\nu<-a$ \,and\, $\nu>2$\,:\\[\jot]
        $f(x)$ has one zero in the region $x<1$ and one in the region $x>1$. In the latter case the mass is positive provided $\eta>L^2\,\sqrt{2+\nu}\,|\alpha|$;}
    \item{$\nu=2$ \,and\, $\eta^2<4\,L^2$\,:\\[\jot]
        $f(x)$ has only one zero in the region $x>1$ and the mass is positive provided $0<\eta<2L$;}
    \item{$\nu=2$ \,and\, $\eta^2>4\,L^2$\,:\\[\jot]
        $f(x)$ has two zeros, one in the region $x<1$ and one in the region $x>1$. Positivity of mass for $x>1$ requires $\eta>2L$;}
    \item{$\nu=-a$\,:\quad
        there exists a zero of $f(x)$ for $x>1$ and the mass is positive there for $\eta>\,L^2 |\alpha|$;}
    \item{$\nu=a$\,:\quad
        there is just one zero of $f(x)$ in the region $x>1$ and the mass is negative.}
    \end{enumerate}
 }
\end{enumerate}

\paragraph{Thermodynamics:}
We can obtain the relevant thermodynamic quantities and show that, as expected, the first law is satisfied without including scalar charge. The hair is `secondary' in the sense that there is no additional integration constant associated with it. Using the usual analytic continuation to the Euclidean section, the temperature of the hairy black hole is given by the periodicity of the Euclidean time, $\beta$, as
\begin{equation}
T\= \frac{1}{\beta}=\frac{L^2}{4\,\pi}\,\frac{x_{+}^{\nu}-1}{\eta\,\nu^2\,x_{+}^{\nu-1}}\,\Big[
(\nu+2)\,\left(\alpha^2 L^2\,(\nu-2)-\eta^2\,k\right)\,x_{+}^{\nu}-(\nu-2)\,
\left(\alpha^2 L^2\,(\nu+2)+\eta^2\,k\right)\Big]\;,
\end{equation}
where $x_{+}$ is the solution of $f(x_{+})=0$.

By a similar rescaling as in (\ref{rho}), the entropy density of the dual field theory becomes
\begin{equation}
s\=\frac{\nu^2\,x_{+}^{\nu-1}}{\eta^2\,\left(x_{+}^{\nu}-1\right)^2}\,\frac{1}{4\,G}\;.
\end{equation}
and one can check that the first law is indeed satisfied:
\begin{equation}
\delta \rho\=T\,\delta s\;.
\end{equation}
Using the mass, entropy, and temperature computed before, we also obtain the free energy density:
\begin{equation}
\label{freeenergy}
F \= M-T\,S \= \left[-\frac{\alpha^2 L^2\,\left(\nu^2-4\right)}{6\,\eta^{3}}
+\frac{k\,\nu (x_{+}^{\nu}+1)}{2 \eta\,\left(x_{+}^{\nu}-1\right)}\right]\,
\frac{\sigma_k L ^4}{8\pi G}\;.
\end{equation}
As an independent check, in Section \ref{euclideanaction}, we are going to compute the regularized action and show that it matches the thermodynamic result (\ref{freeenergy}). We would also like to emphasize that, for  $k=L^{-2}$, the free energy is also bounded from below:
\begin{equation}
F \>
\frac{1}{6\,G}\,\frac{\nu-1}{\left\vert\alpha\right\vert\,\sqrt{\nu-2}}\;.
\end{equation}

\subsubsection{Family 2}
The conformal factor $\Upsilon(x)$ and the ansatz are the same as for the first family (\ref{eq:AdSfunc}). Then, we perform the symmetry transformation \eqref{eq:ST} to find the following new configuration:
\begin{equation}
\begin{split}
\varphi\=&\ell^{-1}\,\ln(x)\;, \qquad\quad\;
\Upsilon(x)~=\frac{L^2 \nu^2\,x^{\nu-1}}{\eta^2\,(x^\nu-1)^2}\;, \\[3.5\jot]
f(x)\=&\frac{x^{2-\nu}\,\left(x^\nu-1\right)^2\,\eta^2\,k}{\nu^2} +
      1+(1-\alpha^2 L^2 )\,\left[-1+\frac{x^2}{\nu^2}\,\left((\nu+2)\,x^{-\nu}-(\nu-2)\,x^\nu+\nu^2-4\right)\right]\;.
\end{split}
\end{equation}
These two families are, in general, not diffeomorphic. In fact, in the limit $\alpha=0$, the horizon disappears for $k=L^{-2}$ and the black hole solutions of Family 1 become naked singularities. However, in the same limit $\alpha=0$, there still exists a horizon for $x>1$ and $\nu>2$, and so the black hole solutions of Family 2 are regular in this limit. We emphasize that the scalar field is now positive when $x>1$ and negative when $x<1$.

\paragraph{Boundary conditions:} Using the change of coordinates (\ref{CC}), we can again explore the boundary conditions of the new hairy black hole solutions. We see that the boundary condition flips its
sign:
\begin{equation}
\label{f2}
\varphi_{1}~=-\frac{\ell}{2}\,\varphi_{0}^2\;,
\end{equation}
while the source is still positive or negative, depending on whether the scalar field is positive or negative. A similar computation for the mass yields the following expression:
\begin{equation}
M=\frac{L^4 \mu_2\,\sigma_k}{8\pi G}\;,\quad\quad
 \quad
\mu_2=\pm\left[\left(1-\alpha^2 L ^2\right)\,\frac{\left(\nu^2-4\right)}{3\,\eta^3}-\frac{k}{\eta}\right]\;.
\end{equation}

\paragraph{Location of the horizon:} The behaviour of $g_{tt}$ around the singularity can be extracted from the analysis of Family 1 and from the symmetry transformation (\ref{eq:ST}). Interestingly, there exist regular hairy black hole solutions for $\varphi_{0}>0$ and
\begin{equation}
\eta^2\,k-(1-\alpha^2 L^{2})\,(\nu-2)\,<\,0 \;,
\end{equation}
which again requires $\nu>2$. In this case, and for spherical topology,
we also found that the existence of a horizon implies that the mass is
positive, with a dual bound:
\begin{equation}
\rho_\text{sph}=\frac{L^2}{8 \pi G} \left[\frac{\nu^2-4}{3\,\eta^3} (1-\alpha^2 L^2)-\frac{1}{\eta L^2}\right]
\;>\;\frac{1}{8 \pi G} \frac{\nu-1}{3\,\eta}\;>\;\frac{1}{24 \pi G  L} \frac{\nu-1}{\sqrt{1-\alpha^2 L^2}  \sqrt{\nu-2}}\;.
\end{equation}
The thermodynamics can be done along the same lines as discussed for Family 1.

\section{Holography of hairy black holes}  \label{sect4:holography}
In this section we interpret the hairy black hole solutions in the context of AdS/CFT duality. We start with a a brief review of mixed boundary conditions for scalar fields and their holographic interpretation. Then, after we obtain the regularized Euclidean action that provides a valid variational principle on the gravity side, we show that the solutions presented in the previous section correspond to RG flows generated by triple-trace deformations of the dual field theory.

\subsection{Dual field theory}
The standard AdS/CFT dictionary identifies the saddle point approximation of
the gravitational action with the large $N$ limit of the partition function of the dual field theory
\begin{equation}
\label{dictionary}
e^{-\frac{L^{2}}{\kappa}\,w(J)}\=
\left\langle e^{N^{3/2}\,\mathcal{C}\int d^{3}x\,J(x)\,\mathcal{O}(x)\,\sqrt{\gamma}}\right\rangle\=
\int\left[d\Phi\right]\,G(\Phi)\,e^{-S\left[\Phi\right]+N^{3/2}\,\mathcal{C}
\int d^{3}x\,J(x)\,\mathcal{O}(x)\,\sqrt{\gamma}}\;,
\end{equation}
where $\kappa = 8\pi G$ and the relation
\begin{equation}
\frac{L^{2}}{\kappa}\=\frac{\sqrt{2}\,N^{3/2}\,K^{1/2}}{12\,\pi}~\equiv~ N^{3/2}\,\mathcal{C}
\end{equation}
has been used, with $K$ the Chern-Simons level of the ABJM theory \cite{Aharony:2008ug, Aharony:2008gk} (see \cite{Billo:2000zr} for an earlier related work).
The term $\frac{L^{2}}{\kappa}\,w(J)$ is the Euclidean gravitational action with the conformal boundary metric representative $\gamma_{\mu\nu}$ and, in the large $N$ limit, $w(J)$, $\mathcal{O}(x)$, and $J(x)$ are of order one.

Consider a single trace operator, $\mathcal{O}$, dual to a bulk field:
\begin{equation}
-\frac{1}{\sqrt{\gamma}}\left.\frac{\delta w(J)}{\delta J}\right\vert _{J=0}\=
\left\langle\mathcal{O}\right\rangle _{J}\;,
\end{equation}
Since the path integral has a saddle point at large $N$, the corrections are suppressed as $1/N$ and so
\begin{equation}
\left\langle\mathcal{O}^{p}\right\rangle \=
\left\langle\mathcal{O}\right\rangle^{p}\;,
\end{equation}
for any $p\in\mathbb{N}$. Using the large $N$ factorization,  whenever the function $W(\mathcal{O})$ has a Taylor expansion around its vacuum expectation value (VEV) $\left\langle\mathcal{O}\right\rangle_{J}=\varphi_{1}$, we have that\note{In general, the right normalization for the CFT VEV is $\left\langle\mathcal{O}\right\rangle_{J}=(\Delta_{+}-\Delta_{-}) \varphi_{1}$, where $\varphi_{1}$ is the coefficient of the faster fall-off branch in the scalar field \cite{Klebanov:1999tb}. All the scalar fields of maximal gauged supergravity have $\Delta_{+}-\Delta_{-}=1$ around the maximally supersymmetric vacuum in four dimensions.}
\begin{equation}
\left\langle W(\mathcal{O})\right\rangle _{J}\=
W(\varphi_{1})+W^{\prime}(\varphi_{1})\big\langle
\left(\mathcal{O}-\varphi_{1}\right)\big\rangle _{J}\;,
\end{equation}
where we have kept the trivial $\left\langle\mathcal{O}-\varphi_{1}\right\rangle_{J}\equiv0$ linear term. We can then rewrite (\ref{dictionary})
\begin{equation}
\left\langle e^{N^{3/2}\,\mathcal{C}
\int d^{3}x\,J(x)\,\mathcal{O}(x)}\right\rangle\=
\left\langle e^{N^{3/2}\,\mathcal{C}
\int d^{3}x\,\left[J\,\mathcal{O}+W(\mathcal{O})-\left(  W(\varphi_{1})+W^{\prime}(\varphi_{1})\,\left(\mathcal{O}-\varphi_{1}\right)  \right)\right]\,\sqrt{\gamma}}\right\rangle\;,
\end{equation}
and, as in \cite{Papadimitriou:2007sj}, we define the deformed current $J_{W}\equiv J-W^{\prime}\left(\varphi_{1}\right)$ to obtain
\begin{equation}
e^{-\frac{L^{2}}{\kappa}\,\bar{w}(J_{W})}\=
e^{-\frac{L^{2}}{\kappa}\,\left[
w(J)-\int d^{3}x\,\left(  W(\varphi_{1})-\varphi_{1}\,W^{\prime}(\varphi_{1})\right)\,
\sqrt{\gamma}\right]}\=
\left\langle e^{N^{3/2}\,\mathcal{C}\int d^{3}x\,\left(
J_{W}\,\mathcal{O}+W(\mathcal{O})\right)\,\sqrt{\gamma}}\right\rangle\;.
\end{equation}

Let us now apply this `holographic prescription' to the case of interest, namely a scalar field with both modes normalizable. It follows that one can
have two possible theories \cite{Klebanov:1999tb}: the standard AdS/CFT prescription associates to the scalar field an operator of conformal dimension $2$, while the alternative theory associates an operator of conformal dimension $1$. In other words, the theories interchange the source $\varphi_{0}$ and the expectation value $\varphi_{1}$, defined by their fall-off in (\ref{fall off}). Therefore, their generating functionals of connected correlation functions must be related by a Legendre transform, which  is analogue to the quantum effective action. Correspondingly, we introduce the notation:
\begin{equation}
\Gamma(\varphi_{1})\=\left[w(\varphi_{0})+\int d^{3}x\,\varphi_{0}\,\varphi_{1}\,\sqrt{\gamma}\right]_{\varphi_{1}\=
-\frac{\delta w}{\delta\varphi_{0}}\frac{1}{\sqrt{\gamma}}}\;,
\end{equation}
where we have used $\left\langle\mathcal{O}\right\rangle _{J}=\varphi_{1}$
and $J=\varphi_{0}$. The effective action, deformed by a function $\bar{W}\left(\varphi_{0}\right)$, is just
\begin{equation}
\Gamma(\bar{J}_{\bar{W}})\=w\left(\varphi_{0}\right)+\int d^{3}x\,
\left[\bar{W}\left(\varphi_{0}\right)-\varphi_{0}\,\bar{W}^{\prime}(\varphi_{0})
+\varphi_{0}\varphi_{1}\right]\,\sqrt{\gamma}\; ,
\end{equation}
where we have used the deformed source $\bar{J}_{\bar{W}}=\varphi_{1}-\bar{W}^{\prime}(\varphi_{0})$. However, it should be stressed that the dictionary is slightly different in this case. Indeed, to identify $\varphi_{0}$ as the VEV and $\varphi_{1}$ is necessary to flip the sign in the exponent of the CFT side of the correspondence (\ref{dictionary})

\begin{equation}
e^{-\frac{L^{2}}{\kappa}\,\Gamma(\bar{J}_{\bar{W}})}\=
\left\langle e^{-N^{3/2}\,\mathcal{C}\int d^{3}x\, (\bar{J}_{\bar{W}} \,\mathcal{O}(x)\,+\bar{W}(\mathcal{O}))\sqrt{\gamma}}\right\rangle\
\end{equation}

\subsection{Euclidean action and variational principle}
\label{euclideanaction}
In this section we closely follow \cite{Anabalon:2015xvl} and consider the  regularized Euclidean action
\begin{equation}
\label{action}
I^\EE\left[g^\EE,\varphi\right]\=
\frac{1}{2\kappa}\int_{M}d^{4}x\,\sqrt{g^\EE}\,\left[  -\frac{R}{2}+\frac{1}{2}\,\left(\partial\varphi\right)^{2}+V(\varphi)\right] - \frac{1}{\kappa}\,\int_{\partial M}d^{3}x\,\sqrt{h}\,K + I_\mathrm{ct} + I_{\varphi}
\end{equation}
that renders the free energy finite and a well posed variational principle. The usual Einstein-dilaton action was supplemented by the gravitational counterterm  $I_\mathrm{ct}$ \cite{Henningson:1998gx, Skenderis:2000in, Balasubramanian:1999re, Mann:1999pc} and scalar field counterterm $I_{\varphi}$ \cite{Papadimitriou:2007sj, Anabalon:2015xvl}:
\begin{equation}
\begin{split}
I_\mathrm{ct}&\=\frac{2}{\kappa}\,\int d^{3}x\,\sqrt{h}\,\left[
    \frac{1}{L}+\frac{L}{4}\,R(h)\right]\;\\
I_{\varphi}&\=\frac{1}{\kappa}\,\int_{\partial M}d^{3}x\,
    \sqrt{h}\,\frac{\varphi^2}{2L}-\frac{L^2}{\kappa}
    \int_{\partial M}d^{3}x\,\sqrt{\gamma}\,\left[
    W(\varphi_{1})-\varphi_{1}\,W^{\prime}(\varphi_1)\right]
\end{split}
\end{equation}
Our conventions are
\begin{equation}
h_{\mu\nu}=g_{\mu\nu}-n_{\mu}n_{\nu}\;, \qquad\;
2K_{\mu\nu}=\mathcal{L}_{n}\,h_{\mu\nu}=\nabla_{\mu}n_{\nu}+\nabla_{\nu}n_{\mu}\;, \qquad\; \gamma_{\mu\nu}=\frac{L^2}{r^2}\,h_{\mu\nu}
\end{equation}
where $n_{\mu}$ is the outwards-pointing normal, $h_{\mu\nu}$ is the induced metric, $K=h_{\mu\nu}\,K^{\mu\nu}$ is the trace of the extrinsic curvature, $L$ is the $AdS_4$ radius, and $\gamma_{\mu\nu}$ is the metric of the geometry where the dual field theory lives. The local counterterm cancels the divergences and its holographic interpretation is standard \cite{Klebanov:1999tb}. The finite counterterm is included as follows from the large $N$ factorization of the previous section.

It is instructive to check that the action (\ref{action}) is stationary under all variations that preserve the mixed boundary conditions of the scalar field. Let us start with the metric ansatz
\begin{equation}
ds^{2}\=N(r)\,dt^{2}+G(r)\,dr^{2}+S(r)\,d\Sigma_{k}\;,
\end{equation}
and the following asymptotic behavior:
\begin{subeqs}
\begin{align}
N(r)&\=\frac{r^2}{L^2}+k\,L^{2}-\frac{\mu\,L^4}{r}+O(r^{-3})\;,\label{fo1}\\
G(r)&\=\frac{L^2}{r^2}-\frac{L^6\,\left(\varphi_0^2+2\,k\right)}{2\,r^4}+\frac{b\,L^8}{r^5}+O(r^{-6})\;,\label{fo2}\\
S(r)&\=\frac{r^2}{L^2}+O(r^{-3})\;,\label{f03}
\end{align}
\end{subeqs}
The asymptotic expansion of the dilaton potential and fall-off of the scalar field are
\begin{equation}
V(\varphi)=-\frac{3}{L^2}-\frac{\varphi^2}{L^2}+O(\varphi^4)
\;, \qquad\;
\varphi\=L^{2}\,\frac{\varphi_{0}}{r}+L^{4}\,\frac{\varphi_{1}}{r^{2}}+O(r^{-3})\;,
\label{fo4}
\end{equation}
where the factors of $L$ have been introduced to match the conformal and
engineering dimensions of $\varphi_{0},\varphi_{1}$, and $\mu$. Thus, the variation of the action yields
\begin{equation}
\delta I^{\EE}=\frac{1}{\kappa}\,\int d^{3}x\,\sqrt{h}\,\delta\varphi\,
    \left(n^\mu\,\partial_\mu \varphi + \frac{\varphi}{L}\right) +\frac{L^{2}}{\kappa}\,\int_{\partial M} d^{3}x\,
    \sqrt{\gamma}\,\varphi_1\,W^{\prime\prime}(\varphi_1)\,\delta\varphi_1\;, \label{varI}
\end{equation}
By using the fall-off (\ref{fo2}) and (\ref{fo4}), we obtain
\begin{subeqs}
\begin{align}
n^{\mu}\,\partial_{\mu}\varphi+\frac{\varphi}{L}&\=
    -L^{3}\,\frac{\varphi_{1}}{r^2}+O(r^{-3})\;,\label{eq1}\\
\delta\varphi&\=L^{2}\,\frac{\delta\varphi_0}{r}+
    L^{4}\,\frac{\delta\varphi_1}{r^{2}}+O(r^{-3})\;.\label{eq2}
\end{align}
\end{subeqs}
We replace $\sqrt{h}=\left(\frac{r}{L}\right)^{3}\sqrt{\gamma}$ and the expressions \eqref{eq1}, \eqref{eq2} in \eqref{varI}, then take the limit for
$r\rightarrow\infty$, and obtain
\begin{equation}
\delta I^{\EE}\=\frac{L^{2}}{\kappa}\,\int d^{3}x\,\sqrt{\gamma}\,
    \left[-\delta\varphi_{0}+W^{\prime\prime}(\varphi_{1})\,\delta\varphi_{1}\right]\,\varphi_{1}
    ~\equiv -\frac{L^{2}}{\kappa}\,\int d^{3}x\,\sqrt{\gamma}\,
    \varphi_1\,\delta J_{W}\;,
\end{equation}
where we have identified the modified source $J_{W}=\varphi_{0}-W^{\prime}(\varphi_{1})$ of the previous section. Hence, the action has an extremum when the source is fixed:
\begin{equation}
\varphi_{0}\=W^{\prime}(\varphi_{1})\;.
\end{equation}
The VEV of the operator with conformal dimension $2$ is given by
$\varphi_{1}$. However, following the standard results of the previous section when the both modes are normalizable, one can also consider $\varphi_{1}$ as the source and $\varphi_{0}$ as the VEV. In this case, the corresponding finite counterterm is
\begin{equation}
+\frac{L^{2}}{\kappa}\,\int_{\partial M}d^{3}x\,\sqrt{\gamma}\,
    \left[\bar{W}(\varphi_{0})-\varphi_{0}\,\bar{W}^{\prime}(\varphi_{0})+
    \varphi_{1}\,\varphi_{0}\right]
\end{equation}
and the variation of the action yields
\begin{equation}
\delta I^{\EE}\=\frac{L^{2}}{\kappa}\,\int d^{3}x\,\sqrt{\gamma}\,
    \left[\delta\varphi_{1}-\bar{W}^{\prime\prime}(\varphi_{0})\,
    \delta\varphi_{0}\right]\,\varphi_{0}~\equiv~
    \frac{L^{2}}{\kappa}\,\int d^{3}x\,\sqrt{\gamma}\,\varphi_{0}\,
    \delta J_{\bar{W}}\;.
\end{equation}
Therefore, in the alternative deformed theory, $\varphi_{0}$ is the VEV and the source is $J_{\bar{W}}=\varphi_{1}-\bar{W}^{\prime}(\varphi_{0})$. The boundary condition is
\begin{equation}
\varphi_{1}~=\bar{W}^{\prime}(\varphi_{0})\;.
\end{equation}
We end this section by pointing out that, as expected, the black hole free energy (\ref{freeenergy}) is related to the regularized Euclidean action by $I^{\EE}\=\beta\,F$, but we do not present the details here.

\subsection{Holographic interpretation}
From the point of view of the field theory, the black holes we induce a triple trace deformation of the operator with conformal
dimension $\Delta = 1$:
\begin{equation}
\varphi_{1}=\pm\frac{\ell}{2}\,\varphi_{0}^2 \quad\;\Longrightarrow\quad\;
\bar{W}(\varphi_0)=\pm\frac{\ell}{6}\,\varphi_{0}^{3}\;,\label{tripletrace}
\end{equation}
where the upper sign corresponds to the first family of hairy black hole solutions, (\ref{f1}), and lower sign to the second one, (\ref{f2}). Let us recall that for spherical and planar black holes, the regularity condition implies that for the first family, $\varphi_{0}<0$, and for the second one, $\varphi_{0}>0$. Therefore, we conclude that the deformation
yields sensible infrared dynamics only when $\bar{W}\left(\varphi_{0}\right)<0$. This boundary condition makes the on-shell Euclidean action of a probe field more negative, as it is shifted by the quantity
\begin{equation}
I_\textsc{cft}\rightarrow I_\textsc{cft}+\frac{L^{2}}{\kappa}\,\int\sqrt{\gamma}\;\bar{W}(\varphi_0)\;,
\label{defaction}
\end{equation}
Since the two possible deformations are mapped to each other by the bulk electromagnetic duality, a self-dual deformation that contains the full space of black hole solution (with the corresponding boundary conditions) can be written as
\begin{equation}
\bar{W}(\varphi_0)\=-\frac{\ell}{6}\,\left\vert\varphi_0\right\vert^3\;.
\end{equation}
The existence of this kind of triple trace deformations, but without the absolute value, was already argued to be necessary to fit the three point functions of ABJM theory at Chern-Simons level $K=1$ \cite{Freedman:2016yue}.

\section{$\mathcal{N}=8$ truncations}  \label{sect5:N8trunc}
In this section we briefly discuss the embedding of our models within maximal four-dimensional supergravity.
The original ${\rm SO}(8)$ gauging of $\mathcal{N}=8,\,D=4$ supergravity \cite{deWit:1981sst,deWit:1982bul} and its generalizations to non-compact/non-semisimple gauge groups ${\rm CSO}(p,q,r)$, $p+q+r=8$, \cite{Hull:1984yy,Hull:1984vg}, were recently shown to be part of a much broader class of gauged maximal theories
\cite{DallAgata:2011aa,DallAgata:2012mfj,DallAgata:2012plb,DallAgata:2014tph,Inverso:2015viq}, also, somewhat improperly, referred to as ``dyonic'' gaugigns, see \cite{Trigiante:2016mnt} for a review. Their construction was effected by exploiting, in the maximal theory, the freedom in the initial choice of the \emph{symplectic frame}, namely of the basis of the symplectic duality representation of the 56 electric and magnetic charges. By gauging, for instance, the same ${\rm SO}(p,q)$ group, $p+q=8$, in different symplectic frames, obtained by rotating the original one of  \cite{deWit:1981sst} by a suitable symplectic matrix, a one-parameter class of inequivalent theories with gauge group ${\rm SO}(p,q)$ were constructed. These are named ${\rm SO}(p,q)_{\omega}$ models, or $\omega$-deformed ${\rm SO}(p,q)$ models, where $\omega$ is the angular variable parametrizing the symplectic frame. Although their string/M-theories origin is as yet obscure, they feature a much richer vacuum structure than their original counterparts \cite{deWit:1981sst,deWit:1982bul,Hull:1984yy,Hull:1984vg} corresponding to the value $\omega=0$ of the angular parameter.\par Global symmetries of the ungauged $\mathcal{N}=8,\,D=4$ supergravity restrict the range of $\omega$ corresponding to inequivalent models:
\begin{align}
\omega&\in \left[0,\,\frac{\pi}{8}\right]\,\,\,,\,\,\,\,\,\mbox{${\rm SO}(8)$ and ${\rm SO}(4,4)$}\,,\nonumber\\
\omega&\in \left[0,\,\frac{\pi}{4}\right]\,\,\,,\,\,\,\,\,\mbox{all other ${\rm SO}(p,q)$ groups}\,.
\end{align}
The infinitely many theories we have described so far contain all the possible one-dilaton
consistent truncations of the $\omega$-deformed ${\rm SO}(8)$ gauged maximal
supergravities. This can seen with the change of variables
\begin{equation}
\alpha\=L^{-1}\sin(\omega)\;.
\end{equation}
The scalar field potential now reads
\begin{equation}
V(\varphi)\=\cos^2(\omega)\;Q(\varphi)+\sin^2(\omega)\;Q(-\varphi)\;,
\end{equation}
where
\begin{equation}
Q(\varphi)\=-\frac{L^{-2}}{\nu^2}\,\left[
\frac{(\nu-1)(\nu-2)}{2}\,e^{-\varphi\,\ell\,(\nu+1)}+2(\nu^{2}-1)\,e^{-\varphi\,\ell}
+\frac{1}{2}\,(\nu+1)(\nu+2)\,e^{\varphi\,\ell\,(\nu-1)}\right]\;.
\end{equation}
Note that now the self-duality invariance of the potential is simply
\begin{equation}
\omega\;\rightarrow\;\omega+\frac{\pi}{2}\;,\qquad\quad
\varphi\;\rightarrow\;-\varphi\;.
\end{equation}
The truncations can be characterized by the breaking of the ${\rm SO}(8)$ gauge group to the following stabilizers of the dilatonic field $\varphi$
\begin{equation}
\begin{tabular}[m]{rcl}%
$\nu=\frac{4}{3}$ & $\rightarrow$ & $\SO(7)$              \;,\\[1ex]
$\nu=2$      & $\rightarrow$ & $\SO(6)\times\SO(2)$  \;,\\[1ex]
$\nu=4$           & $\rightarrow$ & $\SO(5)\times\SO(3)$  \;,\\[1ex]
$\nu=\infty$           & $\rightarrow$ & $\SO(4)\times\SO(4)$  \;.
\end{tabular}
\end{equation}
When \,$\nu=\infty$\, or \,$\nu=2$\, one must also set $\omega=0$ to have an embedding in $\mathcal{N}=8$ supergravity.
This allows to consistently uplift our solutions to corresponding $\omega$-rotated models. Exact examples of these black holes have been analyzed in \cite{Anabalon:2012sn}.

\section{Conclusions}  \label{sect6:discuss}
Since the paper is self-contained with detailed computations and interpretations, we briefly discuss our results and present  possible future directions.

An advantage of working with exact hairy solutions in gravity theories with scalar fields (and developing techniques for constructing new ones) is that it is possible to directly identify some of their generic features. For all exact solutions constructed in this paper, there exists only one integration constant that is related to the black hole mass. As expected, since the hair degrees of freedom are leaving outside the horizon, there is no conserved charge associated to the scalar field and so the hair is `secondary'.

In flat space, it was shown \cite{Gibbons:1996af} that once the boundary conditions are not fixed and the asymptotic value of the dilaton, $\phi_\infty$, can vary, the first law of black hole thermodynamics should be supplemented by a new term $\Sigma\,$d$\phi_\infty$, where $\Sigma$ is the scalar charge. One problem with this proposal is that, in string theory, the scalar fields (moduli) are interpreted as local coupling constants and so a variation of their boundary values is equivalent to changing the couplings of the theory (see \cite{Astefanesei:2006sy, Hajian:2016iyp} for a detailed discussion on scalar charges). In some specific examples (when the boundary conditions break conformal invariance), attempts to write the first law of AdS hairy black hole failed and so one could then ask if there should be a similar term (depending of the scalar charge) in AdS spacetime \cite{Lu:2013ura}.
The resolution of the puzzle was given in \cite{Anabalon:2014fla}, where it was shown that, when the conformal symmetry of the boundary is broken, the Hamiltonian mass has a non-trivial contribution from the scalar field and the first law is satisfied once this contribution is considered.\footnote{The same conclusion can be drawn from \cite{Anabalon:2017eri}, where a concrete relation between the boundary and horizon data was obtained for hairy black holes to show that there is no independent integration constant associated with the scalar field and from \cite{Anabalon:2015xvl} where the boundary conditions that break conformal invariance were studied in detail.}
Since the hairy solutions presented in this paper preserve the conformal symmetry in the boundary, their mass can be easily read-off from the metric \cite{Anabalon:2014fla} and it can be explicitly checked that there is no need to modify the first law by adding scalar charges.

We have obtained two distinct families of black hole solutions that correspond to the static scalar field with triple-trace boundary conditions. In the limit when there is no backreaction, if the boundary CFT is perturbed by an operator of dimension $\Delta$ the corresponding coupling will have dimension $d - \Delta$ and, for our case $\Delta = 1$, the deformation $\bar{W}\left(\varphi_{0}\right)  =\frac{\ell}{6}|\varphi_{0}^{3}|$ will increase the
energy (see, also \cite{Mefford:2014gia}, for another concrete example but for a double trace deformation). It is not clear whether the same conclusion can be drawed in the backreacting case. The backreaction modifies the gravitational contribution to the energy \cite{Anabalon:2016yfg, Anabalon:2017eri}. We leave a more detailed analysis and possible AdS/CMT applications for future work.

We now turn to consider a holographic application and put some emphasis on the physical interpretation of dilaton potential's parameters. Given the holographic connection between the radius in the bulk and energy scale in the boundary theory, we can interpret our black hole solutions as describing RG flows of deformed dual conformal field theory. By considering a variation of the parameter $\nu$, which is a parameter of the theory (not an integration constant of the solution), the deformation should be affected. Since the conformal mass of the dilaton does not change, the conformal dimensions ($\Delta_{\pm} = {2, 1}$) of the operators are invariant, the exact black hole solutions can be interpreted as triple-trace deformations of the dual field theory with the parameter $\nu$ controlling its strength (\ref{tripletrace}), where $\ell^2=2/(\nu^2-1)$. Following \cite{Freedman:1999gp} (see, also, \cite{Goldstein:2005rr} for a similar computation in four dimensions), we use the null energy condition to compute the c-function on the gravity side
\begin{equation}
\mathcal{C}(x)\=
\mathcal{C}_{0}\left[ \frac{x^{\frac{\nu+1}{2}}}{\nu-1+x^{\nu}(\nu+1)}\right]^2
\label{cfuncionBH}
\end{equation}
where $C_0$ is a constant that can be, in principle, fixed by comparing the boundary value, at $x=1$, with the central charge of the dual field theory. This function is non-trivial, except for the following values of the parameter, $\nu = \{-1, 1\}$. As expected, this result fits nicely with the supergravity embedding because the parameter $\nu$ is related to the parameter $n$ in (\ref{modulimetric}) as $n=1+1/\nu$ and so, for $n=\{0,2\}$, the metric of the moduli space vanishes. In this case the flow is trivial and corresponds to the usual Schwarzschild black hole in AdS spacetime.

It is important to emphasize that, unlike even dimensions, a definition of the c-function for a three dimensional field theory is more subtle because the trace of the stress tensor vanishes. However, a proposal for odd dimensions was put forward by Myers and Sinha in \cite{Myers:2010tj}. That is, the central charge is related to the  coefficient of a universal  contribution to the entanglement entropy of the dual field theory. Since we have a concrete embedding in SUGRA for our exact hairy black hole solutions,  it would be interesting to check the result of \cite{Myers:2010tj} in this particular case. Another interesting direction for the future would be to investigate consistent SUGRA truncations when the gauge fields are turned on and to construct exact black hole solutions. In the extremal limit, the near horizon geometry of a static black hole is $AdS_2\times S^2$ and one can apply, for example, the quantum entropy function of Sen \cite{Sen:2008yk, Sen:2008vm} to get the central charge at the IR fixed point. Some similar examples, but in a different context, were recently presented in \cite{Guarino:2017eag, Guarino:2017pkw, Guarino:2017jly} and their holographic microstate counting in \cite{Azzurli:2017kxo, Bobev:2018uxk}.

We have shown that on the solutions we have presented here, regularity implies positivity of the energy. An open question is whether the same is true for all black hole solutions of the theory for these boundary conditions. In the case of double trace deformation and a truncation of type IIB supergravity this was shown to be the case for planar black holes in \cite{Anabalon:2017eri}. It would be interesting to do the same for the maximal supergravity in four dimensions.

Finally, we would like to emphasize that a similar potential can be used in five dimensions to construct exact regular hairy black hole solutions \cite{Acena:2012mr}, though a similar SUGRA construction with FI terms is not possible and it is not clear yet if the scalar field potential can be again obtained from a consistent SUGRA truncation.

\section*{Acknowledgments}

Research of AA is supported in part by Fondecyt Grants 1141073 and 1170279 and Newton-Picarte
Grants DPI20140053 and DPI20140115. The work of DA is supported by the
Fondecyt Grant 1161418 and Newton-Picarte Grant DPI20140115.  We also thank Fondazione CRT \,\includegraphics[height=\fontcharht\font`\B]{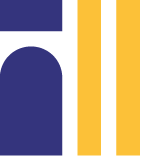}\: that partially supported this work for dott.\ A.\ Gallerati.

%---- APPENDICES ------------------------------------------

\appendix
%\phantomsection % use it for correct TOC link !!!
%\addcontentsline{toc}{section}{Appendices} % add References to TOC
%\addtocontents{toc}{\protect\setcounter{tocdepth}{0}}%no_single_Appendices_in_TOC
%\addtocontents{toc}{\protect\addvspace{2.5pt}}%
\numberwithin{equation}{section}%
\numberwithin{figure}{section}%

%\begin{comment}
%
\section{Lagrangian e matrices}
In this appendix we write the explicit form of the matrices $\II,\,\RR$ in terms of the scalars. The bosonic Lagrangian, with $\kappa=8\pi G=1$, reads
\begin{align}
\frac{1}{\eD}\;\LB&\=
\frac{R}{2}-\frac{1}{2}\,g_{z\bar{z}}\,\dm z\,\dmup
+\frac14\,\II\,\Fd\,\Fu[\Sigma]
+\frac{1}{4}\,\RR\,\Fd\,\hodge\Fu[\Sigma]\,,\\[4\jot]
\hodge\Fd&\=\frac{\eD}{2}\,\LCTd\,\Fu[\Lambda][\rho][\sigma]\;.
\end{align}
\begin{equation}
\RR\=\left(
      \begin{array}{cc}
      \R_{11} & \R_{12} \\
      \R_{21} & \R_{22} \\
      \end{array}
    \right)\;,
\end{equation}
with
\begin{equation}
\begin{split}
\R_{11}&\=\frac{i}{8}\,n\,\z^{-n}\,\zb^{-n}\,\left(\mathscr{A}_{11}-\mathscr{B}_{11}\right)\;,\\[3\jot]
\R_{12}&\=\R_{21}\=\frac{i}{8}\,n\,(n-2)\,
          \Big(\z^{2-n}\,\zb^{n}-\z^{n}\,\zb^{2-n}-(n-1)\left(\z^2-\zb^2\right)\Big) \left(\mathscr{A}_{12}+\mathscr{B}_{12}\right)\;,\\[3\jot]
\R_{22}&\=\frac{i}{8}\,(n-2)\,\z^{-n}\,\zb^{-n}\,\left(\mathscr{A}_{22}-\mathscr{B}_{22}\right)\;,
\end{split}
\end{equation}
and
\begin{equation}
\II\=\left(
      \begin{array}{cc}
      \I_{11} & \I_{12} \\
      \I_{21} & \I_{22} \\
      \end{array}
    \right)\;,
\end{equation}
with
\begin{equation}
\begin{split}
\I_{11}&\=-\frac{1}{8}\,n\,\z^{-n}\,\zb^{-n}\,\left(\mathscr{A}_{11}+\mathscr{B}_{11}\right)\;,\\[3\jot]
\I_{12}&\=\I_{21}\=\frac{1}{8}\,n\,(n-2)\,
          \Big(\z^{n}\,\zb^{2-n}-\z^{2-n}\,\zb^{n}-(n-1)\left(\z^2-\zb^2\right)\Big)
          \left(\mathscr{A}_{12}-\mathscr{B}_{12}\right)\;,\\[3\jot]
\I_{22}&\=-\frac{1}{8}\,(n-2)\,\z^{-n}\,\zb^{-n}\,\left(\mathscr{A}_{22}+\mathscr{B}_{22}\right)\;,
\end{split}
\end{equation}
\begin{equation}
\begin{split}
\mathscr{A}_{11}&\=\frac{2\,\z^{n+2}\,\zb^{n+1}\,\Big((n-2)\,n\,\z-(n-1)^2\,\zb\Big)
                   +(n-2)\,\zb^2\,\z^{2\,n+2}-n\,\z^4\,\zb^{2n}}
                  {\z^n\,\Big((n-2)(n-1)\,\z^2+(n-1)\,n\,\zb^2
                  -2\,n\,(n-2)\,\z\,\zb\Big)+2\,\z^2\,\zb^{n}}\;,\\[3\jot]
\mathscr{A}_{12}&\=\zb\,\bigg(2\,\zb^2\,\z^n+\zb^n\,\Big(n\,(n-1)\,\z^2
                 +(n-2)(n-1)\,\zb^2-2\,n\,(n-2)\,\z\,\zb\Big)\bigg)^{-1}\;,\\[3\jot]
\mathscr{A}_{22}&\=\frac{-2\,\z^{n}\,\zb^{n+1}\,\Big((n-1)^2\,\zb
                 -n\,(n-2)\,\z\Big)+(n-2)\,\zb^2\,\z^{2n}-n\,\z^2\,\zb^{2n}}
                 {2\,\zb^2\,\z^{n}+\zb^{n}\,\Big(
                 n\,(n-1)\,\z^2+(n-2)(n-1)\,\zb^2-2\,n\,(n-2)\,\z\,\zb\Big)}\;,\\[4\jot]
\mathscr{B}_{11}&\=\mathscr{A}_{11}(\z\leftrightarrow\zb)\;,\qquad
\mathscr{B}_{12}\=\mathscr{A}_{12}(\z\leftrightarrow\zb)\;,\qquad
\mathscr{B}_{22}\=\mathscr{A}_{22}(\z\leftrightarrow\zb)\;.
\end{split}
\end{equation}
%
%\end{comment}

%---- BIBLIOGRAPHY ------------------------------------------

\newpage
%\renewcommand{\refname}{\spacedlowsmallcaps{References}} % For modifying the bibliography heading

%\renewcommand{\theequation}{\thesection.\arabic{equation}}
%\numberwithin{equation}{section} % correct number eq in bibliography?

\hypersetup{linkcolor=blue}
\phantomsection % use it for correct TOC link !
\addcontentsline{toc}{section}{References}
\bibliographystyle{myJHEPbibstyle}
\bibliography{gaugedsugra}

\end{document}